\documentclass{emulateapj}
\usepackage{color,amsmath,hyperref,verbatim}
\newcommand{\ie}{i.e.\/}

\newcommand{\etc}{etc.\/}

\shorttitle{Refractive Signal Bending and Delay}
\shortauthors{Mangum \& Wallace}

\begin{document}
\title{Atmospheric Refractive Electromagnetic Wave Bending and Propagation Delay}

\author{Jeffrey G.~Mangum}
\affil{National Radio Astronomy Observatory, 520 Edgemont Road,
  Charlottesville, VA  22903, USA}
\email{jmangum@nrao.edu}

\and

\author{Patrick Wallace}
\affil{RAL Space, STFC Rutherford Appleton Laboratory, Harwell Oxford,
  Didcot, Oxfordshire, OX11 0QX, United Kingdom}
\email{patrick.wallace@stfc.ac.uk}

\begin{abstract}
In this tutorial we summarize the physics and mathematics behind
refractive electromagnetic wave bending and delay.  Refractive bending
and delay through the Earth's atmosphere at both radio/millimetric and
optical/IR wavelengths are discussed, but with most emphasis on the
former, and with Atacama Large Millimeter Array (ALMA) applications in
mind.  As modern astronomical measurements often require sub-arcsecond
position accuracy, care is required when selecting refractive bending
and delay algorithms.  For the spherically-uniform model atmospheres
generally used for all refractive bending and delay algorithms,
positional accuracies $\lesssim 1^{\prime\prime}$ are achievable when
observing at zenith angles $\lesssim 75^\circ$.  A number of
computationally economical approximate methods for atmospheric
refractive bending and delay calculation are presented,
appropriate for astronomical observations under these conditions.  For
observations under more realistic atmospheric conditions, for zenith
angles $\gtrsim 75^\circ$, or when higher positional accuracy is required, 
more rigorous refractive bending and delay algorithms must be
employed.  For accurate calculation of the refractive bending, we 
recommend the \cite{Auer2000} method, using numerical integration to
ray-trace through a two-layer model atmosphere, with an atmospheric
model determination of the atmospheric refractivity.  For the delay
calculation we recommend numerical integration through a model
atmosphere.
\end{abstract}

\keywords{atmospheric effects, telescopes}

\section{Introduction}
\label{Introduction}

The path through the Earth's atmosphere of an electromagnetic
wave emitted by an astronomical source deviates from a
straight line connecting source and observer.  This deviation is due
to changes in the real portion of the refractive index of the Earth's
atmosphere, defined as the ratio of the speed of light in a vacuum and
the phase velocity in the medium through which the electromagnetic wave
propagates:  
\begin{equation}
\label{eq:refractiveindexdef}
n \equiv \frac{c}{v_{phase}}.
\end{equation}
These changes in the refractive index of the atmosphere, combined with
Fermat's principle, which states that an electromagnetic signal will
follow the path between source and observer which takes the least
amount of time, results in a path which is ``curved''.  An observer on
the surface of the Earth measures the effect of this curved path of
the electromagnetic 
signal from the astronomical source as a deflection of the apparent
position of the source and a delay in the arrival time of the
electromagnetic signal.  These two effects are generally referred to
as atmospheric refractive electromagnetic wave bending and delay,
respectively.  Both of these effects lead to errors in astronomical
position measurement.  Atmospheric refractive bending leads to
astronomical position errors measured by single telescopes, while
atmospheric refractive delay leads to position errors measured by
interferometers. 

For refractive signal bending an observer measures a
difference between the unrefracted (or topocentric) zenith
distance\footnote{Astronomers use altitude, elevation ($E$) and zenith
  angle/distance ($z$) interchangeably.  With but one exception, we
  have standardized the analyses presented in this tutorial by using zenith angle.} of
an astronomical source ($z$) and the observed zenith distance ($z_0$) of that source:
\begin{equation}
\label{eq:RefracDef}
R \equiv z - z_0.
\end{equation}
To relate this refractive signal bending to the refractive
  index $n$ we introduce the ``refractivity'' at the observer ($N_0$),
  which is related to refractive index ($n_0$): 
\begin{equation}
\label{eq:refractivitydef}
n_0 - 1 = 10^{-6}N_0,
\end{equation}
where $N_0$, measured in parts per million, is a function of the
atmospheric pressure ($P_0$), temperature ($T_0$), and relative
humidity ($RH_0$) at the observer.

The refractive delay experienced by an incoming electromagnetic wave
due to its propagation through the Earth's atmosphere is given by:
\begin{equation}
\label{eq:DelayDef}
\mathcal{L}_{atm} \equiv \int_s\left(n-1\right)ds
\end{equation}
where $s$ is the path through and $n$ is the refractive index of the
atmosphere.  

The goal of this tutorial is to provide a summary of the standard
models used to calculate atmospheric electromagnetic signal bending
and delay.  With this summary we also discuss the limitations of these
models and their relationship to example astronomical measurements.
All of the refractive models we address are limited by the
simplifications imposed by the parameterization of the Earth's
atmosphere (Section~\ref{Atmosphere}).  For example, all of the
atmospheric models we incorporate in our refractive signal analysis
assume a static, homogeneous, two-layer (troposphere and stratosphere)
atmosphere.  We do not address, for example, effects due to
time-variable atmospheric inhomogeneities (\ie\ scintillation).

In this tutorial we begin by describing the physics of refractive bending
(Section~\ref{Refrac}), which includes a discussion of the
plane-parallel (Section~\ref{PlaneParallel}) and radially-symmetric
(Section~\ref{RadSym}) approximations to the calculation of refractive
bending.  We then review the general formalism used to describe
refractive electromagnetic wave bending through the Earth's atmosphere
(Section~\ref{Bending}), and describe a standard procedure used for
calculating the refractive bending due to the Earth's atmosphere.
This discussion necessarily involves a model of the Earth's atmosphere
(Section~\ref{Atmosphere}).  Our discussion of
atmospheric refractive signal bending ends with a discussion of
commonly-used approximations to the refractive bending
(Section~\ref{ApproximateR}).

Our discussion of refractive delay (Section~\ref{RefDelay}) describes
the general formalism and common usage of ``delay models''.  This
discussion of refractive delay includes an analysis of two additional
corrections to the refractive delay at an antenna which is relevant to
interferometric array observations: differential atmospheric curvature
(Section~\ref{DifferentialCurveDelay}) and antenna height correction
(Section~\ref{DelayHeight}).  Section~\ref{GenFuncReferences} 
provides some background information on some of the generator function
references presented, while Section~\ref{Conclusions} presents our conclusions. 
Throughout this tutorial application of the formalisms presented is
made for the case of the propagation of radio through submillimeter
wavelength electromagnetic radiation.  We use the Atacama Large
Millimeter Array (ALMA) as a source for many of these illustrative
examples.

At a fundamental level the accuracy of the techniques presented in
this tutorial are limited by the simplifications warranted by the need
to model global atmospheric properties using local measurements.
These simplifications include hydrostatic equilibrium for the dry
component 
and uniform mixing of the wet
component (mainly the troposphere) of the Earth's atmosphere.  Another
major source of uncertainty is our limited understanding of the
dispersive and non-dispersive refractive properties of the water
molecule.  We make no attempt to quantify these uncertainties
rigorously, but do provide observational limits to the measured
position of an astronomical source imposed by the simplified
algorithms presented.

\section{The Physics of Refractive Bending}
\label{Refrac}

\begin{figure}
  \centering
  \includegraphics[width=\columnwidth]{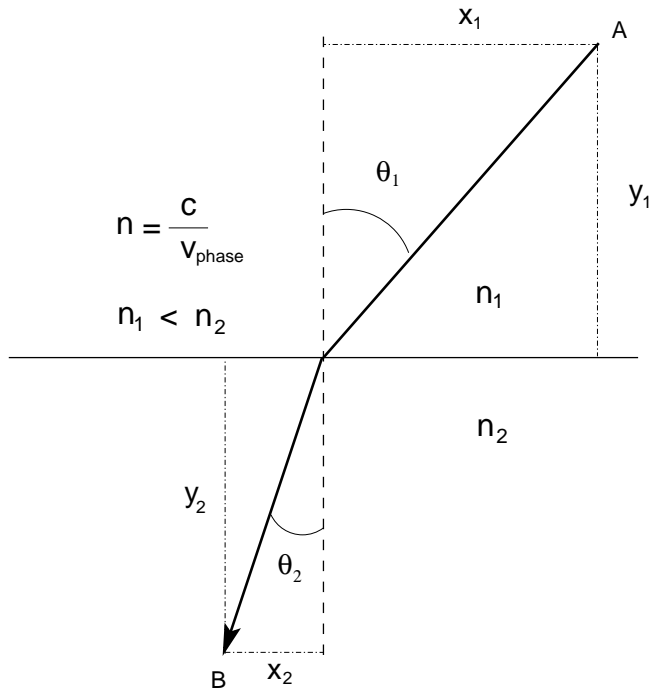}
  \caption{Diagram showing the propagation of an electromagnetic
    signal from one medium to another.} 
  \label{fig:Snell}
\end{figure}

As was noted in Section~\ref{Introduction}, the path of an
electromagnetic wave through a refractive medium, such as the Earth's
atmosphere, is governed by Fermat's Principle.  Figure~\ref{fig:Snell}
displays the example of an 
electromagnetic signal propagating from one medium (\ie\ vacuum) with
index of refraction $n_1$ to another medium (\ie\ the top of the
Earth's atmosphere) with index of refraction $n_2$.  Using the
dimensions illustrated in Figure~\ref{fig:Snell}:
\begin{equation}
\label{eq:Fermat}
t = \frac{\sqrt{x^2_1 + y^2_1}}{v_1} + \frac{\sqrt{x^2_2 + y^2_2}}{v_2},
\end{equation}
where $v_1$ and $v_2$ are the phase velocities of the electromagnetic
signal within each medium.  Using the fact that the total vertical
distance that the electromagnetic signal will travel is given by $d =
y_1 + y_2$, we can substitute for 
$y_2$ in Equation~\ref{eq:Fermat} and differentiate with respect to
$y_1$ in order to find the minimum time needed for the electromagnetic
signal to travel from point A to point B:
\begin{equation}
\label{eq:dtdy}
\frac{dt}{dy_1} = \frac{y_1}{v_1\sqrt{x^2_1 + y^2_1}} - \frac{\left(d-y_1\right)}{v_2\sqrt{x^2_2 + \left(d-y_1\right)^2}}.
\end{equation}
Setting Equation~\ref{eq:dtdy} equal to zero and noting that
$\sin{\theta_1} = y_1/\sqrt{x^2_1 + y^2_1}$,
$\sin{\theta_2} = y_2/\sqrt{x^2_2 + y^2_2}$, and $n =
c/v$, we find that:
\begin{equation}
\label{eq:Snell}
n_1\sin{\theta_1} = n_2\sin{\theta_2},
\end{equation}
which is Snell's Law.  

If we now assume that the refractive medium is
composed of stratified layers which are radially-symmetric about
a common center, we can derive an equation which relates the
total amount of electromagnetic signal refraction to the local atmospheric
conditions at the point of observation.  Before deriving the exact
form for the refraction it is instructive first to derive the
approximate form for the refraction assuming a plane-parallel
atmosphere.  

\subsection{Plane-Parallel Atmosphere}
\label{PlaneParallel}

In the following we derive the approximate form for electromagnetic
signal refraction when the medium through which the signal is
propagating is assumed to be plane-parallel.  This derivation follows
closely and attempts to summarize that presented in three of the
standard references for this work: \cite{Smart1962}, \cite{Bean1966},
and \cite{Green1985}.  
A visualization of a stratified plane-parallel atmosphere is shown in
Figure~\ref{fig:PlaneParallel}.  Consider an atmosphere with $N$
horizontally-stratified layers with refractive indices $n_N$,
$n_{N-1}$, $\ldots$, $n_1$, $n_0$.  An electromagnetic signal entering
the atmosphere at zenith angle $z$ will be successively refracted
through each layer, with the angle of refraction governed by Snell's
Law (Equation~\ref{eq:Snell}):
\begin{figure}
  \centering
  \includegraphics[width=\columnwidth]{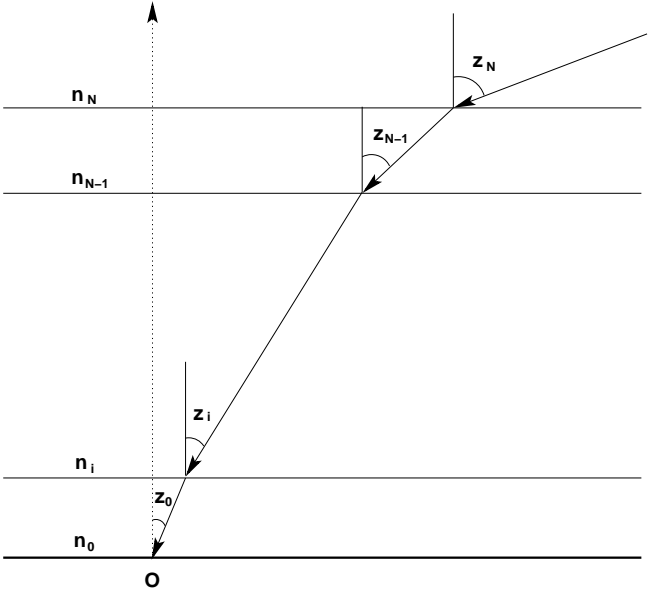}
  \caption{Diagram showing the propagation of an electromagnetic
    signal through a vertically-stratified atmosphere.} 
  \label{fig:PlaneParallel}
\end{figure}
\begin{equation}
\label{eq:Snelliter}
n_i\sin{z_i} = n_{i-1}\sin{z_{i-1}}.
\end{equation}
This successive application of Snell's Law results in the following
relationship between the physical conditions at the top of the
refractive medium and those at the point of observation: 
\begin{eqnarray}
\label{eq:SnellPlane}
n_0\sin{z_0} &=& n_N\sin{z_N} \nonumber \\
&=& \sin{z},
\end{eqnarray}
where we have used the fact that the refractive index of free-space
$n_N$ is 1 and $z_N$ is the unrefracted (or topocentric) zenith
distance $z$.  Defining the angle of refraction $R \equiv z - z_0$,
and noting that $R \ll 1$, we can write Equation~\ref{eq:SnellPlane}
as follows:
\begin{eqnarray}
\label{eq:RPlane}
n_0\sin{z_0} &=& \sin(R + z_0) \nonumber \\
&=& \sin{R}\cos{z_0} + \cos{R}\sin{z_0} \nonumber \\
&\simeq& R\cos{z_0} + \sin{z_0} \nonumber \\
R &\simeq& \left(n_0 - 1\right)\tan{z_0}~\textrm{(radians)}.
\end{eqnarray}
which is the equation for the total refraction in the limit of a
stratified plane-parallel atmosphere.  With the refractivity at the
observer defined by Equation~\ref{eq:refractivitydef}, the refraction
at the observer, $R_0$, is given by:
\begin{equation}
\label{eq:rnotdef}
R_0 = 0.206265 N_0(ppm) \tan{z_o}~\textrm{(arcsec)}.
\end{equation}
Inserting the standard dry atmosphere value for $N_0 \simeq 280$\,ppm
yields:
\begin{equation}
\label{eq:rnotdryatmo}
R_0 \simeq 57.75\tan{z_0}~\textrm{(arcsec)}.
\end{equation}

\subsection{Radially-Symmetric Atmosphere}
\label{RadSym}

In the following we extend the formalism used to derive the refractive
angle induced by a plane-parallel refractive medium to the general
case of a radially-stratified atmosphere
(Figure~\ref{fig:RefractAtmo}).  As with our derivation of the
refraction due to a plane-parallel medium, the following derivation
follows closely and attempts to summarize that presented in three of
the standard references for this work:
\cite{Smart1962}, \cite{Bean1966}, and \cite{Green1985}.  We start
with Snell's Law applied to the first layer of the atmosphere:
\begin{figure}
  \centering
  \includegraphics[width=\columnwidth]{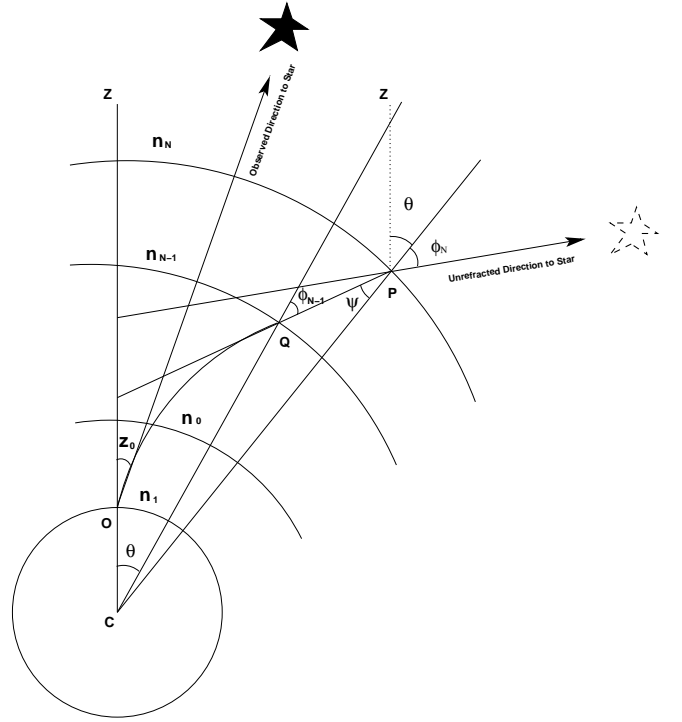}
  \caption{Diagram showing the propagation of an electromagnetic
    signal through a radially-stratified atmosphere.} 
  \label{fig:RefractAtmo}
\end{figure}
\begin{equation}
\label{eq:SnellFirstLayer}
n_N\sin{\phi_N} = n_{N-1}\sin{\psi},
\end{equation}
and noting that, for the triangle CQP, with line segments $CP \equiv
r_N$ and $CQ \equiv r_{N-1}$:
\begin{equation}
\label{eq:TriCQP}
r_{N-1}\sin{\phi_{N-1}} = r_N\sin{\psi}.
\end{equation}
Eliminating $\psi$ from Equations~\ref{eq:SnellFirstLayer} and
\ref{eq:TriCQP}:
\begin{equation}
\label{eq:SnellRadial}
n_N r_N \sin{\phi_N} = n_{N-1} r_{N-1} \sin{\phi_{N-1}}.
\end{equation}
Applying Equation~\ref{eq:SnellRadial} to the last layer of the
atmosphere above the observer:
\begin{equation}
\label{eq:SnellRadialGround}
n r \sin{\phi} = n_0 r_0 \sin{z_0}.
\end{equation}
Noting that the unrefracted (topocentric) zenith angle $z$ is given
by:
\begin{equation}
\label{eq:zdef}
z = \phi_N + \theta,
\end{equation}
and that the angle $\phi_N$ is equal to the angle between $r$ and the
tangent to the angle $\theta$:
\begin{equation}
\label{eq:tanphi}
\tan{\phi_N} = r_N \frac{d\theta}{dr},
\end{equation}
Since it is the variation of $z$ with height above the observer that
produces the total refraction at the observer, we need to take the
differential of Equation~\ref{eq:zdef} and use
Equation~\ref{eq:tanphi}:
\begin{eqnarray}
\label{eq:Diffz}
dz &=& d\phi_N + d\theta \nonumber \\
&=& d\phi_N + \frac{dr}{r_N}\tan{\phi_N}
\end{eqnarray}
and Equation~\ref{eq:SnellRadialGround}:
\begin{eqnarray}
n_0 r_0 \sin{z_0} &=& (n + dn)(r + dr)(\sin{\phi} + d(\sin{\phi}))
\nonumber \\
&=& rn\sin{\phi} + r dn \sin{\phi} + n dr \sin{\phi} +
rn\cos{\phi}d\phi \nonumber
\end{eqnarray}
\begin{equation}
\label{eq:DiffRefrac}
rn\cos{\phi}\left(d\phi + \frac{dr}{r}\tan{\phi}\right) = -r dn \sin{\phi},
\end{equation}
where we have kept only first-order terms in the differentials.
Combining Equations~\ref{eq:Diffz} and \ref{eq:DiffRefrac}: 
\begin{equation}
\label{eq:dz}
dz = -\frac{dn}{n}\tan{\phi},
\end{equation}
after which we can integrate over all layers in the
spherically-symmetric atmosphere, which results in the astronomical
refraction, $R$, defined as the topocentric (\ie\ unrefracted) zenith
angle minus the observed (\ie\ refracted) zenith angle:
\begin{equation}
\label{eq:refrac1}
R = \int_{1}^{n_0}\frac{\tan(z)}{n}dn
\end{equation}
where $n$ is the index of refraction, $z$ is the zenith
angle, and the integral is carried out along the path of the
electromagnetic wave.  In the next section we address the specific
problem of calculating the refractive electromagnetic signal bending
due to the Earth's atmosphere.

\section{Refractive Bending Due to the Earth's Atmosphere}
\label{Bending}

Since the mid-1700s astronomers have studied refractive bending of
electromagnetic waves due to
the Earth's atmosphere in order better to understand the
correspondence between measured and absolute positions of astronomical
objects.  \cite{Young2004} presents a very thorough historical review
of the development of our understanding of atmospheric refraction at
optical wavelengths.  The development of radio refraction algorithms
parallels that described by \cite{Condon2004}  for the Green Bank
Telescope.  There have been many formulations of the
equation which describes the bending of light through the Earth's
atmosphere \citep[see][]{Young2004}.  The following derivation of a
generalized refractive bending calculation using a simple ray-trace
analysis was 
originally proposed by \cite{Auer1979}\footnote{Although
  \cite{Young2000} reports that the algorithm had in fact been derived
  and used by J.~B.~Biot in 1839.} and further developed by 
\cite{Hohenkerk1985}, and is described in \cite{Urban2013}.  A
modern description of the algorithm can be found in \cite{Auer2000}.
The \textsl{SLALIB}\footnote{\textsl{SLALIB} is the name of a widely used
collection of positional-astronomy computer subprograms.  A Fortran
version released under the GNU General Public License is available
from the Starlink Software Store:  see
http://starlink.jach.hawaii.edu/starlink.  Proprietary C versions
exist also.} refraction function \textsl{slaRefro} uses a modified 
version of the \cite{Hohenkerk1985} development of the \cite{Auer1979}
algorithm.  Recent versions of \textsl{slaRefro} include an atmospheric model
\citep{Liebe1993} that allow for calculation of the atmospheric
refractivity up to frequencies of 1\,THz.

In principle, the refraction $R$ could be calculated directly from
Equation~\ref{eq:refrac1} by numerical quadrature.  But, as
\cite{Auer1979,Auer2000} point out, numerical difficulties at $z=90^\circ$
make it preferable to use $z$ itself as the variable of integration.
\cite{Auer2000} derive a transformed version of
Equation~\ref{eq:refrac1} which varies slowly over $z$ and avoids the  
numerical difficulties at $z=90^\circ$.  Following their derivation,
Equation~\ref{eq:refrac1} can be written in terms of $\ln(n)$ as follows:
\begin{equation}
\label{eq:refrac2}
R = \int_{0}^{\ln(n_0)}{\tan{z}}~d(\ln{n}).
\end{equation}
Taking the logarithmic derivative of Equation~\ref{eq:SnellRadialGround}
\begin{eqnarray}
\label{eq:snell2}
\ln(rn) &=& \ln(n_0r_0\sin{z_0})-\ln(\sin{z}) \nonumber \\
\frac{d(\ln(rn))}{dz} &=& -\frac{1}{\tan{z}}
\end{eqnarray}
and substituting this expression into Equation~\ref{eq:refrac2}
\begin{equation}
\label{eq:refrac3}
R = -\int_{0}^{\ln(n_0)}\frac{dz}{d(\ln(rn))}d(\ln{n}).
\end{equation}
Further substituting the following
\begin{eqnarray}
\label{eq:expansion1}
d(\ln(rn)) &=& d(\ln{r}) + d(\ln{n}) \nonumber \\
R(\ln{n_0}) &=& R(z_0)
\end{eqnarray}
leads to
\begin{eqnarray}
\label{eq:refrac4}
R &=& -\int_{0}^{z_0}\frac{d(\ln{n})}{d(\ln{r}) + d(\ln{n})}dz
  \nonumber \\
  &=& -\int_{0}^{z_0}\frac{\frac{d(\ln{n})}{d(\ln{r})}}{1 + \frac{d(\ln{n})}{d(\ln{r})}}dz,
\end{eqnarray}
which is Equation~3 from \cite{Auer2000}.  Making the substitution
\begin{equation}
\label{eq:expansion2}
\frac{d(\ln{n})}{d(\ln{r})} = \frac{r}{n}\frac{dn}{dr}
\end{equation}
leads to the following
\begin{equation}
\label{eq:refrac5}
R = -\int_{0}^{z_0}\frac{r\frac{dn}{dr}}{n + r\frac{dn}{dr}}dz.
\end{equation}
Note that one can replace the refractive index $n$ with the
refractivity $N$ using Equation~\ref{eq:refractivitydef}.

Equation~\ref{eq:refrac5} is well-behaved at $z = 90^\circ$ and can be
evaluated by quadrature using equal steps in $z$.  At each step in $z$
the corresponding values for $r$, $n$, and $\frac{dn}{dr}$ must also
be calculated, thus requiring input from a model of the radial
variation of P, T, and RH in the Earth's atmosphere (see
Section~\ref{Atmosphere}).  Values for $r$, $n$, and $\frac{dn}{dr}$
are found by finding the roots of Equation~\ref{eq:SnellRadialGround}
as a function of $r$:
\begin{equation}
\label{eq:Fofr}
F(r) = n r - \frac{n_0 r_0 \sin{z_0}}{\sin{z}}.
\end{equation}
One can find the root of Equation~\ref{eq:Fofr} by Newton-Raphson
iteration, whereby the following equation is calculated with an
initial guess ($r_0$) to find successive potential roots ($r_1$,
$r_2$, \ldots):
\begin{eqnarray}
\label{eq:ri}
r_{i+1} &=& r_i - \frac{F(r_i)}{F^\prime(r_i)} \nonumber \\
       &=& r_i - \Biggl[\frac{n_i r_i - n_0 r_0
  \frac{\sin{z_0}}{\sin{z}}}{n_i + r_i\frac{dn_i}{dr_i}}\Biggr]
\end{eqnarray}
for $i = 1, 2, \ldots$, where $r_i$ is the value of r calculated at
the previous step of the integration, and we have used the fact that
\begin{equation}
\label{eq:fprime}
F^\prime(r) = \frac{dn}{dr}r + n.
\end{equation}
Convergence of this iteration is fast, requiring only about 4 steps.
Once one has a converged solution for $r$, $n$ and $\frac{dn}{dr}$ can
be calculated using the chosen atmospheric model.

The calculation then continues by integrating Equation
\ref{eq:refrac5} over each atmospheric interval (troposphere and
stratosphere) using Simpson's rule with summation over equal steps in $z$
\begin{equation}
\label{eq:simpsons}
\int^{r_3}_{r_0}f(r)dr =
\frac{\Delta r}{3}\left(f_0+4f_1+2f_2+f_3\right),
\end{equation}
where $f_n$ is $f(x)$ evaluated at $x = x_0, x_1, x_2,\textrm{ and } x_3$.
One can then compare each integration result with the result of the
previous step of this integration.  There is then a check for either
convergence (\textsl{slaRefro} uses $|\int f(z_i)dz-\int f(z_{i-1})dz| \leq
10^{-8}$) or maximum iteration reached (\textsl{slaRefro} uses 16384).
If convergence or maximum iteration has not been reached, recalculate
$r$ at each step in zenith distance by again solving
Equation~\ref{eq:refrac5} using the procedure outlined above
(Equation~\ref{eq:ri}).

Equation~\ref{eq:refrac5} is the refraction equation used in
\cite{Urban2013}, Equation~7.80.  A simple two component model of the
atmosphere is often assumed.  In this model, there is a discontinuity
in $\frac{dn}{dr}$ at the tropopause, so the refraction integral must
be calculated in two parts: one for the troposphere and another for
the stratosphere.  Note also that atmospheric inhomogeneities can be
accounted for in this formalism by using multiple components in the
integration.

\subsection{Atmospheric Model}
\label{Atmosphere}

Equation \ref{eq:refrac5} requires a description of the radial
variation of $n$ and its derivative $\frac{dn}{dr}$, which depend upon
the radial variation of $P$, $T$, and $RH$ in the Earth's atmosphere.
A number of analytic expressions for $n(r)$ and $\frac{dn}{dr}$ have
been used in the past, including the piecewise polytropic model of
\cite{Garfinkel1944,Garfinkel1967}.  Following the atmospheric model
described by \cite{Sinclair1982} and \cite{Hohenkerk1985}, a simple
two-component model for the Earth's atmosphere can be defined as follows: 
\begin{itemize}
\item Spherically symmetric distribution of density with two layers
  (troposphere and stratosphere).
\item Hydrostatic equilibrium.
\item Perfect gas law applies.
\item Temperature decreasing at a constant rate with height in the
  troposphere and constant in the stratosphere.
\item The Gladstone-Dale relation, $n-1=a\rho$, which relates the
  refractive index $n$ and the density $\rho$, where $a$ is a
  constant which depends only on the local physical properties of the
  atmosphere.
\item Two layer structure\footnote{In the adopted atmospheric model
  the tropopause is a transition, not a layer.} with $a < \infty$ for
  $r_e \leq r \leq h_t$ and $a = \infty$ for $h_t \leq r \leq h_s$.
\item Constant relative humidity in the troposphere which is
  consequently equal to the relative humidity measured at the
  observer.
\item The following constants:
   \begin{itemize}
   \item Universal gas constant:\\ $R_g = 8314.32~J/(mole*K)$
   \item Molecular weight of dry air:\\ $M_d = 28.9644~gm/mole$
   \item Molecular weight of wet air:\\ $M_w = 18.0152~gm/mole$
   \item Molecular weight of atmosphere (mixture of dry and wet air): $M_{atm}$
   \item Acceleration due to gravity at the center of mass of the
     vertical column of air above the observer at observer height
     $h_0$: $g_m$.  See Appendix~\ref{gm} for further details on the
     preferred expression for $g_m$.
   \item Height of the Earth's geoid (assuming WGS84 spheroid) as a
     function of latitude: $r_{WGS84} =
     6378.137\left(1-\frac{\sin^2{\phi}}{298.257223563}\right)~km$ 
   \item Distance from the geoid to the observer: $h_0$
   \item Distance from the geoid to the tropopause: $h_t$
   \item Distance from the geoid to the limit of the stratosphere: $h_s$
   \item Total height of the observer: $r_0 = r_{WGS84} + h_0$
   \item Total height of the troposphere:\\ $r_t = r_{WGS84} + h_t$
   \item Total height of the stratosphere:\\ $r_s = r_{WGS84} + h_s$
   \end{itemize}
\end{itemize}

In the following we derive the radial variation of the temperature ($T$)
and pressure ($P$).

\subsubsection{Temperature Distribution}
\label{Temperature}

The distribution of temperature with $r$ is defined as:
\begin{eqnarray}
\label{eq:tdist}
T(r) &=& T_0+\alpha(r-r_0) \nonumber \\
\frac{dT}{dr} &=& \alpha,
\end{eqnarray}
where $\alpha$ is often referred to as the ``atmospheric temperature
lapse rate''.  In the following analysis of the pressure distribution we
will use these temperature relations.

\subsubsection{Pressure Distribution}
\label{Pressure}

In the following we derive the distribution of pressure with height
above the observer.  The algorithm we describe follows closely that
presented by \cite{Sinclair1982}, \cite{Murray1983}, and
\cite{Hohenkerk1985}.  Combining the ideal gas law: 
\begin{equation}
\label{eq:ideal}
P = \frac{\rho R_g T}{M_{atm}}
\end{equation}
and the equation for hydrostatic equilibrium:
\begin{equation}
\label{eq:hydrostat}
\frac{dP}{dr} = -g_m\rho
\end{equation}
and the temperature distribution relation
(Equation~\ref{eq:tdist}) we find that:
\begin{equation}
\label{eq:dpoverp}
\frac{dP}{P} = -\frac{g_m M_{atm}}{\alpha R_g}\frac{dT}{T}.
\end{equation}
Integrating Equation~\ref{eq:dpoverp} yields:
\begin{eqnarray}
\label{eq:intdpoverp}
\int\frac{dP}{P} &=& -\frac{g_m M_{atm}}{\alpha R_g}\int\frac{dT}{T}
\nonumber \\
\ln\left(\frac{P}{P_0}\right) &=&
\ln\left(\frac{T}{T_0}\right)^{-\frac{g_m M_{atm}}{\alpha R_g}} \nonumber \\
\frac{P}{P_0} &=&
\left(\frac{T}{T_0}\right)^{-\frac{g_m M_{atm}}{\alpha R_g}} \nonumber \\
             &=& \left(\frac{T}{T_0}\right)^\beta
\end{eqnarray}
where we have defined:
\begin{equation}
\label{eq:beta}
\beta \equiv -\frac{g_m M_{atm}}{\alpha R_g}.
\end{equation}

The total atmospheric pressure ($P$) and density ($\rho$) each have
two components: the partial pressure and density due to dry air
($P_d$ and $\rho_d$) and the partial pressure and density due to water
($P_w$ and $\rho_w$). Since the water vapor pressure $P_w$
decreases much more rapidly than the total pressure $P$, we need to
separate $P$ into its constituent parts.  These pressures and
densities are related as follows:
\begin{eqnarray}
\label{eq:pandrho}
P &=& P_d + P_w \\
\rho &=& \rho_d + \rho_w
\end{eqnarray}
using the Ideal Gas Law (Equation~\ref{eq:ideal}) for each
component (dry, wet, and total), we can write Equation~\ref{eq:ideal}
as:
\begin{eqnarray}
\label{eq:ideal2}
P &=& \frac{R_g T}{M_{atm}}\left(\rho_d + \rho_w\right) \nonumber \\
  &=& \frac{P_d M_d + P_w M_w}{M_{atm}},
\end{eqnarray}
which allows us to write $M_{atm}$ in terms of its dry and
wet components as using Equation~\ref{eq:pandrho}:
\begin{eqnarray}
\label{eq:m}
M_{atm} &=& \frac{P_d M_d + P_w M_w}{P} \nonumber \\
       &=& M_d - \frac{P_w\left(M_d - M_w\right)}{P}.
\end{eqnarray}
Combining Equations~\ref{eq:m}, \ref{eq:dpoverp}, and
\ref{eq:intdpoverp} produces a general expression which describes the
variation of $P$ with $r$:
\begin{eqnarray}
\label{eq:dpoverp2}
\frac{dP}{P} &=& \frac{-g_m M_d}{\alpha R_g}\frac{dT}{T} + \frac{g_m M_d
  P_w}{\alpha R_g
  P_0}\left(\frac{T}{T_0}\right)^{-\beta}\left(1 -
  \frac{M_w}{M_d}\right)\frac{dT}{T} \nonumber \\ 
 &=& \beta\frac{dT}{T} -
 \beta\frac{P_w}{P_0}\left(\frac{T}{T_0}\right)^{-\beta}\left(1
   - \frac{M_w}{M_d}\right)\frac{dT}{T}.
\end{eqnarray}
Note that in Equation~\ref{eq:dpoverp2} $g_m$
(Equation~\ref{eq:gmwgs84}) and $T$ (Equation~\ref{eq:tdist}) are
known functions of $r$.  Only the radial dependence of $P_w$ is as yet
unknown.

At this point we need to take a little diversion into the relationship
between relative humidity ($RH$) and saturation vapor pressure
($e_{sat}$).  In Appendix~\ref{RHandPsat} we note that the
approximation:
\begin{equation}
\label{eq:psatdep4}
\frac{e_{sat}(P,T)}{e_{sat}(P_0,T_0)} = \left(\frac{T}{T_0}\right)^\gamma
\end{equation}
for saturation vapor pressure agrees with the more exact
expression (Equation~\ref{eq:rhesat}: \cite{Buck1981}) to within
$\pm0.2$\,hPa\footnote{Note that 1 hectopascal (hPa) = 1 millibar
  (mb) and that we use these two units interchangeably.} for P between
600\,hPa and 1200\,hPa and T between $-30$\,C and $+20$\,C.
Therefore, using Equation~\ref{eq:psatdep4} in 
Equation~\ref{eq:dpoverp2} yields:
\begin{equation}
\label{eq:dpoverp3}
\frac{dP}{P} = \beta\frac{dT}{T} -
 \beta\frac{P_{w0}}{P_0}\left(\frac{T}{T_0}\right)^{\gamma-\beta}\left(1
   - \frac{M_w}{M_d}\right)\frac{dT}{T}.
\end{equation}
Integrating Equation~\ref{eq:dpoverp3} in the same way as for
Equation~\ref{eq:dpoverp} leads to the general expression 
which describes the radial dependence of atmospheric pressure:
\begin{eqnarray}
\label{eq:intdpoverp3}
\ln\left(\frac{P}{P_0}\right) &=& \ln\left(\frac{T}{T_0}\right)^\beta +
\frac{\beta}{\gamma-\beta}\left(1 -
  \frac{M_w}{M_d}\right) \nonumber \\
  &&\frac{P_{w0}}{P_0}\left[1 -
  \left(\frac{T}{T_0}\right)^{\gamma-\beta}\right] \nonumber \\
\frac{P}{P_0} &=& \left(\frac{T}{T_0}\right)^\beta \exp(W)
\end{eqnarray}
where we have defined:
\begin{equation}
\label{eq:w}
W \equiv \frac{\beta}{\gamma-\beta}\left(1 -
  \frac{M_w}{M_d}\right)\frac{P_{w0}}{P_0}\left[1 -
  \left(\frac{T}{T_0}\right)^{\gamma-\beta}\right].
\end{equation}
\cite{Sinclair1982} points out that $W \lesssim 0.003$,
which allows one to expand the exponential as $\exp(W) \simeq 1 + W$
and write Equation~\ref{eq:intdpoverp3} as:
\begin{multline}
\label{eq:intdpoverp4}
\frac{P}{P_0} = \left(\frac{T}{T_0}\right)^\beta +
\frac{\beta}{\gamma-\beta}\left(1 -
  \frac{M_w}{M_d}\right) \\
  \frac{P_{w0}}{P_0}\left[\left(\frac{T}{T_0}\right)^\beta
  - \left(\frac{T}{T_0}\right)^\gamma\right].
\end{multline}

\subsubsection{Application to the Troposphere and Stratosphere}
\label{Application}

In the following we list the parametric forms for $P(r)$, $T(r)$,
$RH(r)$, $n$, and $\frac{dn}{dr}$ in the troposphere and the stratosphere:
\begin{description}
\item[Troposphere:] ($r_e \leq r \leq h_t$)
\begin{eqnarray}
\label{eq:dentrop}
T(r) &=& T_0+\alpha(r-r_0) \\
P(r) &=& P_0\left(\frac{T}{T_0}\right)^\beta +
\frac{\beta P_{w0}}{\gamma-\beta}\left(1 -
  \frac{M_w}{M_d}\right) \nonumber \\
  &&\left[\left(\frac{T}{T_0}\right)^\beta
  - \left(\frac{T}{T_0}\right)^\gamma\right] \\
RH(r) &=& RH_0~\textrm{(constant)} \\
n &=& 1 + 10^{-6}N(r) \\
\frac{dn}{dr} &=& 10^{-6}\frac{dN(r)}{dr}
\end{eqnarray}
\item[Stratosphere:] ($h_t \leq r \leq h_s$)

For isothermal atmospheric layers (like the stratosphere), $\alpha =
0$ and we use the approximation $\ln(1+\epsilon)\rightarrow\epsilon$
as $\epsilon\rightarrow 0$, which makes Equations~\ref{eq:tdist} and
\ref{eq:intdpoverp} become
\begin{eqnarray}
\label{eq:denstrato}
T(r) &=& T(h_t) \textrm{~(constant)} \\
P(r) &=& P(h_t)\exp\left[-\frac{g_m M_{atm}(r-r_t)}{R_gT(h_t)}\right] \\
RH(r) &=& 0 \\
n &=& 1 + (n(h_t)-1)\exp\left[-\frac{g_m
    M_{atm}(r-r_t)}{R_gT(h_t)}\right] \nonumber \\
  &=& 1 + 10^{-6}N(h_t)\exp\left[-\frac{g_m M_{atm}(r-r_t)}{R_gT(h_t)}\right] \\
\frac{dn}{dr} &=&
-\frac{g_m M_{atm}(r-r_t)}{R_gT(h_t)}(n(r_t)-1) \nonumber \\
  && \exp\left[-\frac{g_m M_{atm}(r-r_t)}{R_gT(r_t)}\right] \nonumber \\
  &=& 
-\frac{g_m M_{atm}(r-r_t)}{R_gT(h_t)}10^{-6}N(r_t) \nonumber \\
  && \exp\left[-\frac{g_m M_{atm}(r-r_t)}{R_gT(r_t)}\right] 
\end{eqnarray}
\end{description}

\subsubsection{Atmospheric Radio/Submillimeter Refractivity}
\label{N0rad}

There are two ways to derive the atmospheric refractivity $N_0$ at the
observatory for use in Equation~\ref{eq:refrac5}:
\begin{enumerate}
\item Develop a closed-form expression for $N^{rad}_0$ as functions of $P$ and
  $T$.
\item Use an atmospheric model.
\end{enumerate}
As the historical development of $N_0$ started with (1), which will also
allow us to describe the physics behind this quantity, we revisit
those expressions for $N_0 \equiv N^{rad}_0$ which are appropriate for calculations at
radio and submillimeter wavelengths\footnote{For a brief description
  of atmospheric refractivity at optical wavelengths, see Appendix
  \ref{N0opt}}.

In general, the refractivity of moist air at microwave frequencies
depends upon the permanent and induced dipole 
moments of the molecular species that make up the atmosphere.  The
primary species that make up the dry atmosphere, nitrogen and oxygen,
do not have permanent dipole moments, so contribute to the
refractivity via their induced dipole moments.  Water vapor does have
a permanent dipole moment.  Permanent dipole moments contribute to the
refractivity as $N^{rad}_0\propto \frac{P}{T^2}$, while induced dipole moments
contribute as $N^{rad}_0\propto \frac{P}{T}$, where $P$ is the pressure and
$T$ is the temperature of the species. 

A simple parameterization of the frequency-independent (nondispersive)
refractivity at the zenith is given by the Smith-Weintraub equation
\citep{Smith1953}:
\begin{equation}
\label{eq:SmithWeintraub}
N^{rad}_0 = k_1\frac{P_d}{T}+k_2\frac{P_w}{T}+k_3\frac{P_w}{T^2}+k_4\frac{P_c}{T}
\end{equation}
where $P_d$, $P_w$, and $P_c$ are the partial pressures due to dry
air, water vapor, and carbon dioxide, T is the temperature of the
atmosphere, and $k_1$, 
$k_2$, $k_3$, and $k_4$ are constants.  The dry and wet air
refractivities are then given by:
\begin{eqnarray}
\label{eq:zdzw}
N_d &=& k_1\frac{P_d}{T} \\
N_w &=& k_2\frac{P_w}{T}+k_3\frac{P_w}{T^2} \\
N_c &=& k_4\frac{P_c}{T} = \frac{5}{3}\frac{P_c}{T}.
\end{eqnarray}
Since the partial pressure due to carbon dioxide is $\sim
0.04\%$\footnote{At present, this compares with less than 0.03\% in
pre-industrial times, and is currently increasing by more than
0.002\% per decade.} of
the total pressure, this term is often ignored or lumped into the dry
air contribution in the simple parameterizations of atmospheric
refractivity.

The dry air contribution to this refractivity ($N_d$) is primarily due
to oxygen and nitrogen, and is nearly in hydrostatic equilibrium.
Therefore, $N_d$ does not depend upon the detailed behavior of dry
air pressure and temperature along the path through the atmosphere,
and can be derived based on local atmospheric temperature and pressure
measurements.  The wet air refractivity ($N_w$) can be inferred from
local water vapor radiometry measurements.  

Closed-form approximations for the nondispersive $N^{rad}_0(P,T)$ have been
derived for use at frequencies below 100 GHz by \cite{Brussaard1995}:
\begin{flalign}
\label{eq:nobw}
  ^{BW}N^{rad}_0 &= 77.6\frac{P_d}{T} + 72.0\frac{P_w}{T} + 3.75\times 10^5
  \frac{P_w}{T^2}~~ppm \nonumber \\
 &= 77.6\frac{P}{T} - 5.6\frac{P_w}{T} + 3.75\times 10^5
  \frac{P_w}{T^2}~~ppm 
\end{flalign}
and \cite{Smith1953} (see also \cite{Crane1976}
and \cite{Liebe1977}):
\begin{flalign}
\label{eq:nosw}
  ^{SW}N^{rad}_0 &= 77.6\frac{P_d}{T} + 72.0\frac{P_w}{T} +
3.776\times 10^5 \frac{P_w}{T^2}~~ppm \nonumber \\
 &= 77.6\frac{P}{T} - 12.8\frac{P_w}{T} +
3.776\times 10^5 \frac{P_w}{T^2}~~ppm 
\end{flalign}
where
\begin{itemize}
  \item[$P_d$] is the partial pressure of dry gases in the atmosphere
(in hPa),
  \item[$P_w$] is the partial pressure of water vapor (in hPa),
  \item[$P$] is the total barometric pressure (in hPa), which is equal
    to $P_d + P_w$, and
  \item[$T$] is the ambient air temperature (in Kelvin).
\end{itemize}

The best of the closed-form approximations to the nondispersive
refractivity, though, is the equation derived by \cite{Rueger2002}
which uses what he describes as the ``best average'' values for the
coefficients $k_1$, $k_2$, and $k_3$ (which includes a 375 ppm
contribution due to carbon dioxide in the $k_1$ term):
\begin{flalign}
\label{eq:norba}
  ^{Rueger}N^{rad}_0 &= 77.6890\frac{P_d}{T} + 71.2952\frac{P_w}{T}
& \nonumber \\
 & \qquad\qquad + 3.75463\times 10^5 \frac{P_w}{T^2}~~ppm \nonumber \\
 &= 77.6890\frac{P}{T} - 6.3938\frac{P_w}{T} & \nonumber \\
 & \qquad\qquad + 3.75463\times 10^5 \frac{P_w}{T^2}~~ppm.
\end{flalign}

\begin{figure*}
  \centering
  \includegraphics[trim=15mm 5mm 10mm 20mm, clip, angle=-90, width=\textwidth]{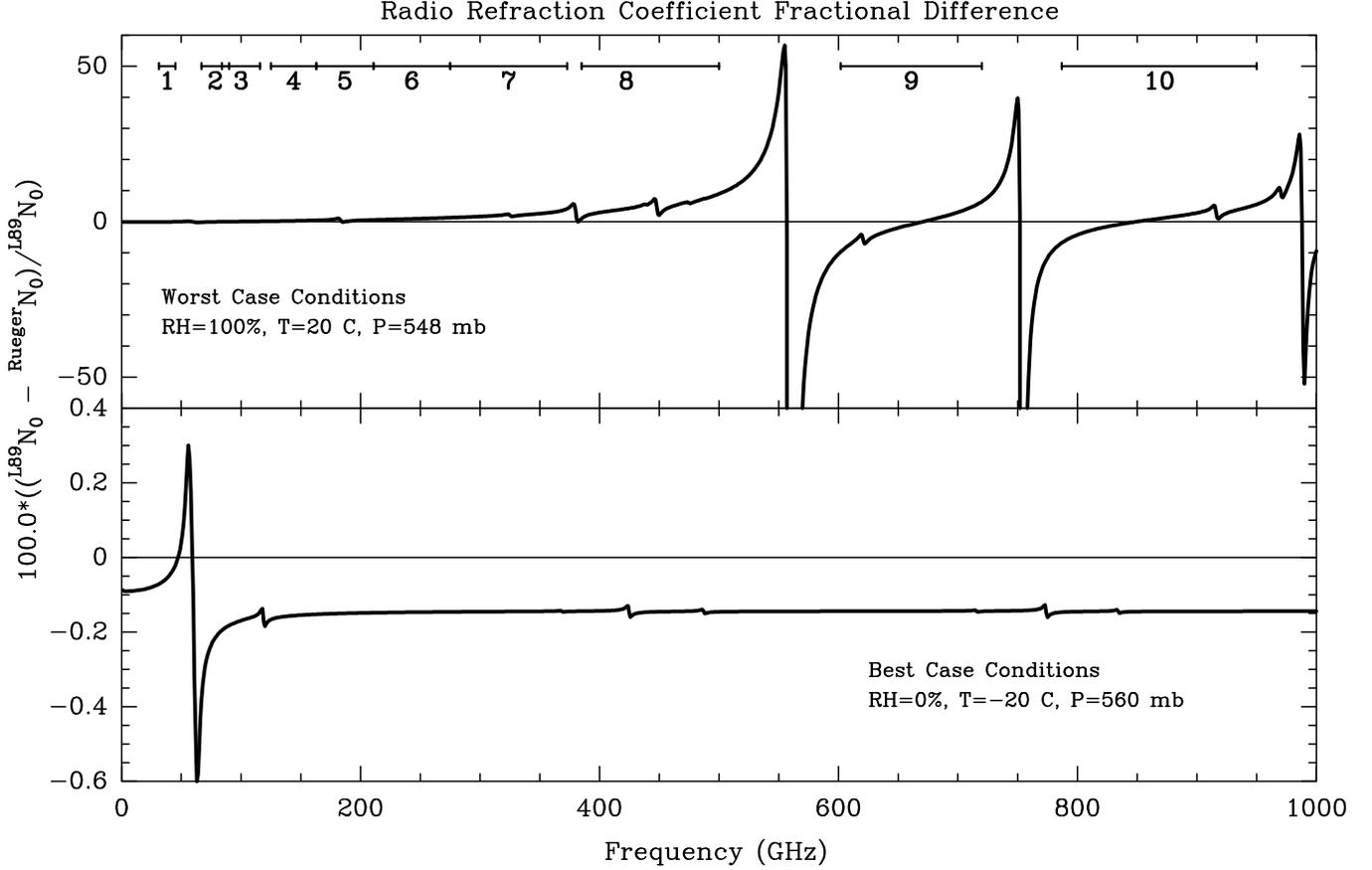}
  \caption{Radio refraction coefficient fractional difference between
    the \cite{Rueger2002} and \cite{Liebe1989} estimates for
    $N^{rad}_0$ under the worst (top panel) and best (bottom panel)
    sets of atmospheric conditions measured at the ALMA site. The
    best-case condition is equivalent to a troposphere devoid of water
    vapor.  The solid horizontal bars near the top of the diagram show the frequency
    ranges for the 10 ALMA receiver bands (\cite{Wootten2009}).} 
  \label{fig:RefractivityComp}
\end{figure*}

Comparing these three closed-form expressions for radio refractivity
at representative values of pressure, temperature, and relative
humidity appropriate for the best (P = 560\,hPa, T = $-20$\,C, RH =
0\%) and worst (P = 548\,hPa, T = $+20$\,C, RH = 100\%) atmospheric
conditions at the ALMA site (altitude = 5.0587\,km) to a more exact model
of the atmospheric refractivity (which includes a dispersive
contribution), we find that:
\begin{itemize}
  \item The \cite{Brussaard1995}, \cite{Smith1953}, and
    \cite{Rueger2002} expressions agree to better than 0.1\% for
    all conditions. 
  \item The \cite{Brussaard1995}, \cite{Smith1953}, and
    \cite{Rueger2002} expressions agree with a more exact
    (\ie\ including dispersive refractivity; \cite{Liebe1989})
    atmospheric model prediction of $N^{rad}_0$ to better than (see
    Figure~\ref{fig:RefractivityComp}):
  \begin{itemize}
    \item Under the best ALMA atmospheric conditions:
    \begin{itemize}
      \item 0.08\% at 8\,GHz
      \item 0.13\% at 230\,GHz
      \item 0.13\% at 370\,GHz (this is a band edge for ALMA)
      \item 0.13\% at 950\,GHz (the highest band edge for ALMA)
    \end{itemize}
    \item Under the worst ALMA atmospheric conditions:
    \begin{itemize}
      \item 0.11\% at 8\,GHz
      \item 0.76\% at 230\,GHz
      \item 3.85\% at 370\,GHz
      \item 6.42\% at 950\,GHz
    \end{itemize}
  \end{itemize}
\end{itemize}
It is clear from this comparison that the closed-form expressions
for $N_0$ are very good for calculations at frequencies far from
telluric lines and for relatively dry conditions.  For general
high-accuracy calculations at submillimeter wavelengths which require
better than 5\% accuracy one should use
an atmospheric model \citep[such as][]{Liebe1989,Liebe1993,Pardo2001}
which incorporates both nondispersive and dispersive contributions to
the refractivity to derive the total atmospheric refractivity.

\subsection{Approximations to the Astronomical Refraction}
\label{ApproximateR}

Instead of using the integral Equation \ref{eq:refrac1}, various
approximations are often made to reduce this expression to a simple
analytic form.  Some of the more generally useful forms are based on
a generator function formalism which assumes an exponential
atmospheric profile
\begin{equation}
\label{eq:atmo}
N(h) = N_0 \exp\left[-\frac{(r-r_0)}{H}\right],
\end{equation}
where $r$ and $r_0$ are height coordinates and $H$ is the
effective height of the atmosphere
\begin{equation}
\label{eq:H}
H = \frac{R_gT}{M_{atm}g_m},
\end{equation}
where $R_g$ is the universal gas constant, $M_{atm}$ is the molar
mass of the atmosphere, $T$ is the temperature of the atmosphere, and
$g_m$ is the gravitational acceleration constant measured at the
center of the vertical column of air (see Section~\ref{Atmosphere}).

One form of this generator function formalism has been described by
\cite{Yan1995} and \cite{Yan1996} as follows: 
\begin{equation}
\label{eq:refrac-yan}
R_{generator} = R_0 m^{\prime}(z)\sin{z},
\end{equation}
where $R_0$ is defined in terms of $N_0$ in
Equation~\ref{eq:rnotdef} and where the generator function
$m^\prime(z)$ is defined as follows:
\begin{equation}
\label{eq:mprime}
m^{\prime}(z) = \cfrac{1}{\cos{z} + \cfrac{A_1}{I^2\sec{z} +
\cfrac{A_2}{\cos{z} + \cfrac{13.24969}{I^2\sec{z} + 173.4233}}}}
\end{equation}
with
\begin{equation}
\label{eq:I}
I = \sqrt{\frac{r_0}{2H}} \cot{z}.
\end{equation}
See \cite{ALMAmemo366} for further information on the use of
this formalism for calculating the refraction.  Note, though, that the
analysis presented in \cite{Yan1995} purports to achieve an accuracy far
better than is realistic.  Furthermore, comparisons with the
refraction function \textsl{slaRefro} suggests that the parametric
equation presented in \cite{Yan1995} is tuned to a specific set of
site and meteorological conditions (sea level and relatively dry).

\begin{figure*}
  \centering
  \includegraphics[trim=10mm 10mm 10mm 50mm, clip, angle=-90, width=\textwidth]{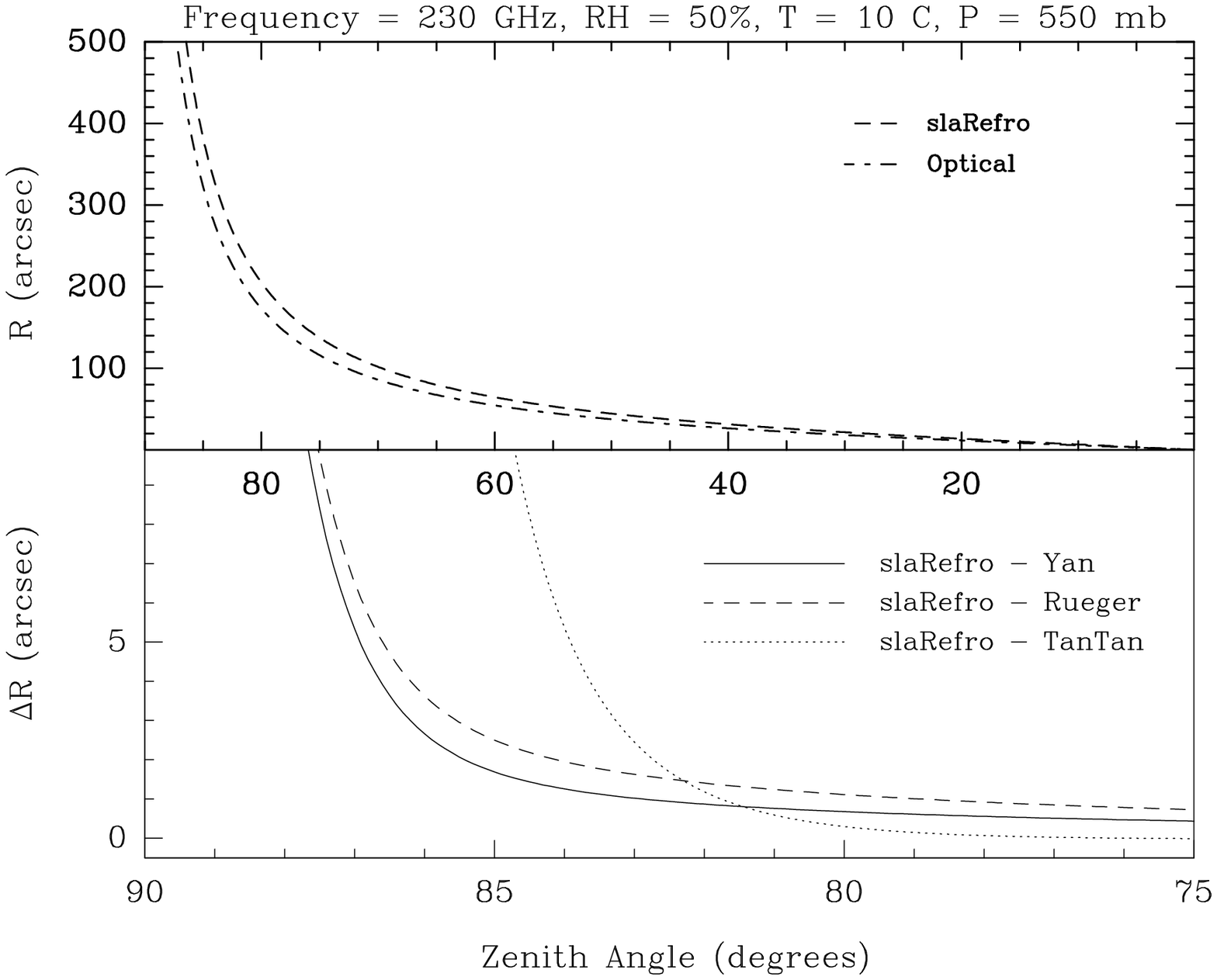} \\
  \caption{Refraction (R; top) and refraction difference ($\Delta$R;
    bottom) as a function of zenith angle for a sampling of refraction
    models.  The refraction function \textsl{slaRefro} is a 
    modified version of the \cite{Hohenkerk1985} development of the
    \cite{Auer1979} algorithm (Section~\ref{Bending}).  ``Optical'' uses
    Equation~\ref{eq:nopt} with the \cite{Yan1995} generator function
    Equation~\ref{eq:refrac-yan}. ``Rueger'' uses
    Equation~\ref{eq:norba} with with the \cite{Yan1995} generator
    function Equation~\ref{eq:refrac-yan}.  ``TanTan'' uses
    Equation~\ref{eq:RsphericalApprox} with the coefficient C set to
    zero and coefficients A and B derived using the \textsl{SLALIB}
    routine \textsl{slaRefco}.  To derive A and B \textsl{slaRefco}
    forces the refraction Equation~\ref{eq:RsphericalApprox} to agree
    with \textsl{slaRefro} at $z = 45^\circ$ and $\arctan(4)$, or $\sim
    76^\circ$.}
  \label{fig:RefracExamples}
\end{figure*}

An even simpler, though less exact, approximation to
Equation~\ref{eq:refrac1} can be derived if one assumes a single-layer
uniform atmosphere.  Noting that Snell's Law (Equation~\ref{eq:Snell})
reformulated in terms of zenith angle for a single-layer atmosphere
(Equation~\ref{eq:SnellRadialGround} with $\phi = z$) becomes:
\begin{equation}
\label{eq:SnellRadialGroundZ}
n r \sin{z} = n_0 r_0 \sin{z_0},
\end{equation}
we can solve for $\sin{z}$ and substitute into the trigonometric
identity for $\tan{z}$: 
\begin{eqnarray}
\label{eq:tanz}
\tan{z} &=& \frac{\sin{z}}{\sqrt{1-\sin^2{z}}} \nonumber \\
&=& \frac{n_0 r_0 \sin{z_0}}{\sqrt{n^2 r^2 - n^2_0 r^2_0 \sin^2{z_0}}}.
\end{eqnarray}
Substituting this expression into our general equation for atmospheric
refraction (Equation~\ref{eq:refrac1}) results in an approximation to
the refractive atmospheric bending due to a single-layer Earth atmosphere:
\begin{equation}
\label{eq:Rspherical}
R_{spherical} = \int_{1}^{n_0}\frac{n_0 r_0 \sin{z_0}}{n(n^2 r^2 - n^2_0 r^2_0
\sin^2{z_0})^{\frac{1}{2}}}dn.
\end{equation}
As noted in \citet[][ Chapter III, Section~ 37]{Smart1962},
since the height of the Earth's atmosphere at which the refractive
medium is located is small in comparison with its radius ($r \ll r_0$),
we can use:
\begin{equation}
\label{eq:roverrnot}
\frac{r}{r_0} = 1 + \epsilon
\end{equation}
where $\epsilon \ll 1$ to substitute for $\frac{r}{r_0}$ in
Equation~\ref{eq:Rspherical}:
\begin{multline}
\label{eq:RsphericalExpand}
R_{spherical} = \int_{1}^{n_0}\frac{n_0 \sin{z_0}}{n(n^2 - n^2_0
\sin^2{z})^{\frac{1}{2}}}dn \\
- \int_{1}^{n_0}\frac{n_0 \sin{z_0} n \epsilon}{(n^2 - n^2_0
\sin^2{z})^{\frac{3}{2}}}dn,
\end{multline}
which after integration \cite[See][]{Smart1962,Green1985} becomes: 
\begin{equation}
\label{eq:RsphericalApprox}
R_{spherical} = A\tan{z} + B\tan^3{z} + C\tan^5{z}
\end{equation}
where A, B, and C are constants dependent on the local
atmospheric temperature, pressure, and relative humidity.  The
approximations used to derive Equation \ref{eq:RsphericalApprox} are
good for $z \leq 75^\circ$.  For a plane-parallel single-layer
atmosphere all of the terms higher than first order in $z$ are zero,
which results in the following equation for the atmospheric refraction
(see Equation~\ref{eq:RPlane}):
\begin{equation}
\label{eq:Rplane}
R_{plane} = A\tan(z).
\end{equation}
Figure~\ref{fig:RefracExamples} shows some example refraction
calculations.

\section{Refractive Delay Due to the Earth's Atmosphere}
\label{RefDelay}

The calculation of the atmospheric refractive delay parallels that for
refractive bending.  To illustrate this fact, the plane-parallel
atmosphere approximation to the general equation for atmospheric delay
(Equation~\ref{eq:DelayDef}) is given by:
\begin{equation}
\label{eq:DelayPlane}
\mathcal{L}_{atm} = \int^\infty_{r_0}\frac{10^{-6}N(r)}{\cos{z}}dr
\end{equation}
In practice the upper limit to the integral in
Equation~\ref{eq:DelayPlane} is the top of the stratosphere.  By using an
atmospheric model to calculate $N(r)$ one can numerically integrate
Equation~\ref{eq:DelayPlane} to derive the refractive delay due to the
atmosphere.  Note that Equation~\ref{eq:DelayPlane} becomes inaccurate
at large zenith angles.

To derive a more exact estimate of the atmospheric refractive delay
one can assume an atmosphere that is horizontally stratified with an
exponential distribution in scale height.  \cite{TMS2001} pp.~516-518
discuss this scenario, the derivation for which we reproduce in the
following.  The excess path length is given by:  
\begin{equation}
\label{eq:diffdelay}
\mathcal{L}_{atm} = 10^{-6} N_0 \int_0^\infty\exp\left(-\frac{h}{h_{atm}}\right) dy,
\end{equation}
where $N_0$ is the refractivity at the Earth's surface, $h$
is the height above the Earth's surface, $h_{atm}$ is the atmospheric
scale height, $y$ is the length coordinate along the direction to the
source, $z$ is the antenna zenith angle while observing the source,
and an exponential distribution to the atmospheric index of refraction
has been assumed.  One can
relate $y$, $h$, $h_{atm}$, and $z$ as follows (see Figure 13-4 in
\cite{TMS2001}, page 517) using the cosine rule on the triangle formed
by $r_0$, $y$, and $r_0+h$:
\begin{equation}
\label{eq:rnoth}
\left(r_0 + h\right)^2 = r^2_0 + y^2_0 - 2r_0y\cos(180^\circ-z).
\end{equation}
Solving for $h$ yields:
\begin{equation}
\label{eq:h}
h = y\cos{z} + \frac{y^2 - h^2}{2r_0}.
\end{equation}
For the nearly right-angled triangle with sides
$y\sin(z_i)$, $y$, and $h$, we can write: 
\begin{equation}
\label{eq:ysquared}
y^2 - h^2 \simeq \left(y\sin{z_i}\right)^2.
\end{equation}

\begin{figure*}
  \centering
  \includegraphics[angle=-90, width=\textwidth]{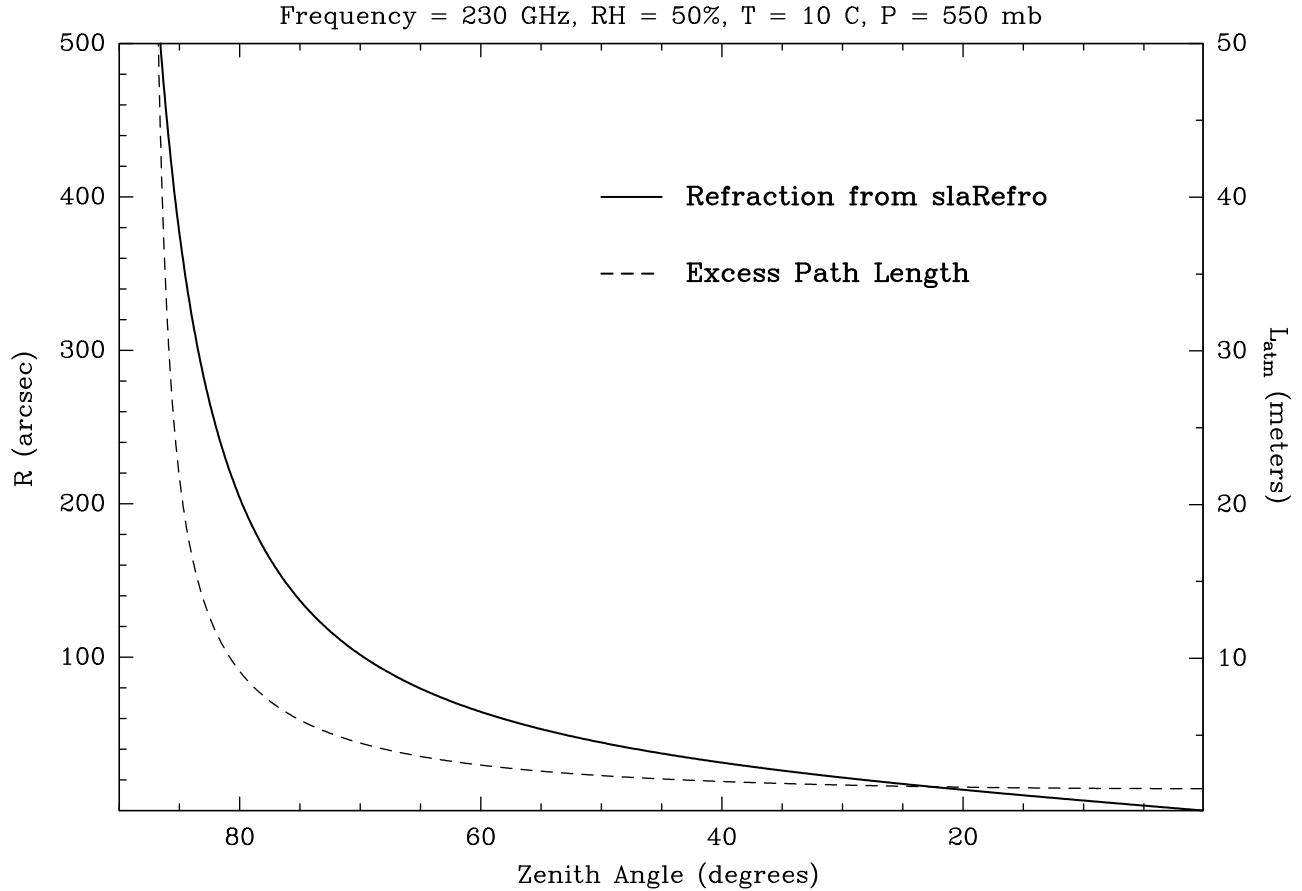}
  \caption{Refraction (R; left) and excess path delay
    ($\mathcal{L}_{atm}$; right) as a function of zenith angle for the
    atmospheric conditions indicated.  The \cite{Liebe1989}
    atmospheric model has been used to calculate $\mathcal{L}_{atm}$
    ($N_0 = 189.416$\,ppm).}
  \label{fig:RefDelayCalcPlot}
\end{figure*}

Since $r_0\simeq 6370$\,km and $h\simeq 12$\,km (the typical height of
the troposphere, which varies from 9 to 17\,km, pole to equator, and
seasonally), $r_0 \gg h$.  Since $z_i \simeq z+\frac{h}{r_0}$,  $z_i 
\simeq z$ (refractive bending is neglected).  The equation for $h$ in
terms of $y$, $z$, and $r_0$ then becomes: 
\begin{equation}
\label{eq:h2}
h \simeq y\cos{z} + \frac{y^2}{2r_0}\sin^2{z}.
\end{equation}
We can now write the expression for $\mathcal{L}$ as follows:
\begin{multline}
\label{eq:diffdelay2}
\mathcal{L}_{atm} \simeq 10^{-6} N_0
\int_0^\infty\exp\left(-\frac{y}{h_{atm}}\cos{z}\right) \\
\times\exp\left(-\frac{y^2}{2r_0h_{atm}}\sin^2{z}\right) dy.
\end{multline}
Since $\frac{y^2}{r_0h_{atm}} \ll 1$, the second term in the
equation above can be expanded with a Taylor series so that: 
\begin{multline}
\label{eq:diffdelay3}
\mathcal{L}_{atm} \simeq 10^{-6} N_0
\int_0^\infty\exp\left(-\frac{y}{h_{atm}}\cos{z}\right) \\
\times\left(1-\frac{y^2}{2r_0h_{atm}}\sin^2{z}+\frac{y^4}{8r^2_0h^2_0}\sin^4{z}+...\right)
dy.
\end{multline}
Integration yields:
\begin{multline}
\label{eq:intdiffdelay}
\mathcal{L}_{atm} \simeq 10^{-6} N_0 h_{atm}\sec{z} \\
\times\left(1-\frac{h_{atm}}{r_0}\tan^2{z}+\frac{3h^2_0}{r^2_0}\tan^4{z}+...\right).
\end{multline}
Writing this equation in terms involving $\sec{z}$, the
excess path length $\mathcal{L}$ becomes: 
\begin{multline}
\label{eq:intdiffdelay2}
\mathcal{L}_{atm} \simeq 10^{-6}N_0h_{atm} 
\Biggl[\left(1+\frac{h_{atm}}{r_0}+\frac{3h^2_0}{r^2_0}\right)\sec{z} \\
-\left(\frac{h_{atm}}{r_0}+\frac{6h^2_0}{r^2_0}\right)\sec^3{z}+\frac{3h^2_0}{r^2_0}\sec^5{z}+...\Biggr].
\end{multline}
Note that one must calculate $N_0$ using a suitable atmospheric model
which uses measurements of the local atmospheric pressure, temperature
and relative humidity to derive the resultant differential residual
delay.  In Figure~\ref{fig:RefDelayCalcPlot} we show a representative
calculation of $\mathcal{L}_{atm}$ for the same set of typical
atmospheric conditions on the ALMA site used to characterize the
atmospheric refractive bending ($R$) in Figure~\ref{fig:RefracExamples}.

\subsection{Differential Excess Atmospheric Delay Between Antennas}
\label{DifferentialDelay}

In the calculation of the atmospheric delay for the antenna elements
in an interferometer, two additional delay corrections need to be
considered.  The first correction is due to the differential excess
path length induced by a non-planar atmosphere \citep{Hinder1971}.
The second is a correction to $N_0$ at each antenna which accounts for
differences in the height of the antenna (Az,El) intersection point
above the reference point for the local atmospheric parameters for the
interferometer. 

\subsubsection{Differential Atmospheric Curvature Delay Between
  Antennas}
\label{DifferentialCurveDelay}

\begin{figure}
  \centering
  \includegraphics[width=\columnwidth]{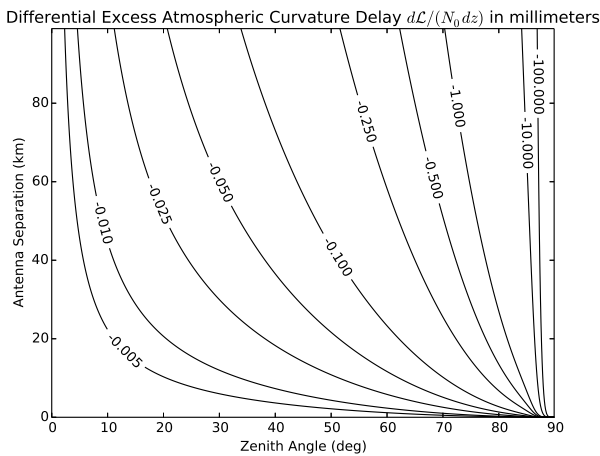}
  \caption{Plot of Equation~\ref{eq:diffcurvedelay} as a function of
    $N_0$ for baseline lengths from 10\,m to 1\,km and
    zenith angle 1 to 90 degrees.}
  \label{fig:diffcurvedelaycalc}
\end{figure}

For antenna elements oriented along an east-west baseline observing a
source that is transiting, we can estimate the change in excess
atmospheric delay between one antenna and another antenna along this
baseline.  Taking the derivative of $\mathcal{L}_{atm}$ with respect
to $z$ and multiplying this derivative by the baseline length $D$
divided by $r_0$ yields the atmospheric differential delay between two
antennas separated by distance $D$ along an east-west baseline:
\begin{multline}
\label{eq:diffcurvedelay}
\frac{d\mathcal{L}_{atm}}{dz} \simeq \frac{-DN_0h_0\tan{z}}{r_0}
\Biggl[\left(1+\frac{h_0}{r_0}+\frac{3h^2_0}{r^2_0}\right)\sec{z} \\
-3\left(\frac{h_0}{r_0}+\frac{6h^2_0}{r^2_0}\right)\sec^3{z}+\frac{15h^2_0}{r^2_0}\sec^5{z}+...\Biggr]~~(mm)
\end{multline}
where $D$ is in m, $h_0$ is in km, $r_0$ is in km, and the result is
in mm.  Figure~\ref{fig:diffcurvedelaycalc} shows the results of
Equation~\ref{eq:diffcurvedelay} as a function of $N_0$ for a range of
baseline lengths and source zenith angles.  To illustrate the magnitude
of this correction to the atmospheric delay, for an antenna separation
of $\sim 2$\,km observing a source at a zenith angle of
$\sim45$\,degrees the differential excess atmospheric curvature delay
is $\sim 5N_0$\,$\mu$m.  For a typical value of $N_0\sim300$\,ppm
$\frac{d\mathcal{L}_{atm}}{dz} \simeq 1.5$\,mm.

\subsubsection{Antenna Height Correction to Total Atmospheric Delay}
\label{DelayHeight}

In the calculation of the zenith atmospheric delay at an antenna it is
assumed that the atmospheric properties (P, T, RH) are the values
measured at the (Az,El) axis intersection point of the antenna.  For
example, in VLBI each antenna station has a set of associated weather
measurements which are used to calculate $N_0$.  For a clustered array
like the VLA or ALMA, the effects of the differences in antenna
(Az,El) axis intersection point height above some reference point for
the local atmospheric parameters need to be accounted for.

The antenna height correction to the total atmospheric delay can be
estimated using a simple atmospheric delay model which corrects for
the path difference between each antenna in an array and a reference
point at the center of the array.  For a clustered array like the VLA,
the extra atmospheric path due to a difference in antenna height above
the center-of-the-array reference point ($\Delta H$, in ns) is given
simply by the change in atmospheric pressure between the antenna array
elements.  A simple estimate of the magnitude of the antenna height
difference correction at the zenith can be obtained by assuming that
the pressure P changes linearly with height.  Then, for example,
100\,cm of additional antenna height out of a total atmospheric height
of 8\,km would correspond to $\left(\frac{100~cm}{8~km}\right) P =
0.099$\,hPa of pressure differential, where we have assumed that P =
790\,hPa (typical atmospheric pressure at the VLA site).  The change
to the dry term of the atmospheric delay is roughly 2.3\,mm/hPa.  This
implies that a pressure change of 0.099\,hPa corresponds to
approximately $0.228$\,mm of path difference, which is approximately
ten times smaller than the excess delay due to atmospheric curvature
(Section~\ref{DifferentialCurveDelay}).

\subsection{Refractive Delay Calculation in Practice}
\label{RefDelayCalc}

Starting in the 1970's geodesists developed atmospheric refractive
delay models which emphasized computational simplicity.  As with the
derivation of the atmospheric refractive bending, atmospheric
refractive delay is generally parameterized as the product of a term
which depends upon the local atmospheric parameters ($Z$) and a term
which describes the zenith angle dependence of the atmospheric delay
through the use of a ``mapping function''\footnote{The ``mapping
  function'' $M$ used in atmospheric refractive delay calculations is
  directly analogous to the ``generator function'' $m^\prime$ sometimes
  used in the atmospheric refractive bending calculation.} ($M$).

As was the case for atmospheric refractive bending, in lieu of an
atmospheric model based calculation of $N$ it is often convenient to
separate the atmospheric delay into contributions due to the dry and
wet components of the atmosphere:
\begin{equation}
\label{eq:DelayATM}
\mathcal{L}_{atm} = \mathcal{L}_d+\mathcal{L}_w,
\end{equation}
where $\mathcal{L}_d$ is the contribution due to dry air while $\mathcal{L}_w$ is
the contribution due to water vapor.  In general $\mathcal{L}_d$ and $\mathcal{L}_w$ are
parameterized in terms of a zenith contribution to the delay which is
dependent upon local atmospheric conditions ($Z$) and a mapping
function ($M$) which relate delays at an arbitrary zenith angle
$z$ to that at the zenith:
\begin{eqnarray}
\label{eq:DelayATM2}
\mathcal{L}_{atm} &=& ZM \nonumber \\
           &=& Z_d M_d + Z_w M_w
\end{eqnarray}

Since $z$ is the unrefracted zenith angle, refractive delay effects
are included in the mapping functions $M$.  In the following we
describe typical methods for calculating $Z$ and $M$.

\subsubsection{Zenith Delay}
\label{ZenithDelay}

The contribution to the atmospheric delay at the zenith ($Z$) is a
measure of the integrated refractivity of the atmosphere at the
zenith ($N$).  As was noted in Section~\ref{N0rad} there are closed-form
expressions for $N(P,T)$ which are appropriate for calculations at
frequencies below 100 GHz.  For high-frequency calculations, one must
use an atmospheric model.

\subsubsection{Mapping Functions}
\label{MappingFunction}

The simplest form for the mapping function ($M$), which relates the
delay at an arbitrary zenith angle $z$ to that at the zenith, is
given by the plane-parallel approximation for the Earth's atmosphere: 
\begin{equation}
\label{eq:mapping1}
M = \frac{1}{\cos{z}}
\end{equation}
This simple form is in practice inadequate, which led
\cite{Marini1972} to consider corrections which accounted
for the Earth's curvature.  Assuming an exponential atmospheric
profile where the atmospheric refractivity varies exponentially with
height above the antenna, \cite{Marini1972} developed a continued
fraction form for the mapping function: 
\begin{equation}
\label{eq:Marini}
M =
\cfrac{1}{\cos{z}+\cfrac{a}{\cos{z}+\cfrac{b}{\cos{z}+c}}},
\end{equation}
where we include only the first three terms in the continued
fraction.  The constants $a, b, c, d$, \etc\ in the continued fraction
forms for the mapping functions presented in this tutorial are
generally derived from analytic fits to ray-tracing results of
standard atmospheric models.   These mapping function constants are
often optimized using measurements of the atmospheric distribution of
pressure and temperature over an observatory (based on
radiosonde measurements, for example).  The mapping functions derived in
\cite{Niell1996} and \cite{Davis1985} are optimized in this way.

Two slight modifications to the
\cite{Marini1972} continued fraction functional form can be
implemented to force $M=1$ at the zenith:
\begin{itemize}
\item Normalize Equation~\ref{eq:Marini} as follows:
\begin{equation}
\label{eq:MariniNormal}
M =
\cfrac{1+\cfrac{a}{1+\cfrac{b}{1+c}}}{\cos{z}+\cfrac{a}{\cos{z}+\cfrac{b}{\cos{z}+c}}}.
\end{equation}
See \cite{Niell1996} for a discussion of how to use this form of the
mapping function\footnote{Note that Equation 4 in \cite{Niell1996}
  contains a typo.  The numerator should be just $A$, rather than
  $\frac{1}{A}$.  See \cite{Niell2001}.  Equation~\ref{eq:MariniNormal}
  lists the correct form for this equation.}, including derivation of the
coefficients $a$, $b$, and $c$.
\item Replace the even numbered $\cos{z}$ terms (\ie\ the second,
  fourth, sixth, etc.) with $\cot{z}$:
\begin{equation}
\label{eq:Chao}
M=\cfrac{1}{\cos{z}+\cfrac{a}{\cot{z}+\cfrac{b}{\cos{z}+c}}}.
\end{equation}
\cite{Chao1974} introduced this modification by truncating
the \cite{Marini1972} form to include only two terms.
\end{itemize}
As noted in Section~\ref{ApproximateR}, a similar continued-fractional
form for the mapping function has been developed by \cite{Yan1995} and
\cite{Yan1996} (Equation~\ref{eq:mprime}).

A physically more correct mapping function has been derived by
\cite{Lanyi1984}.  Unlike previous mapping functions, Lanyi's does not
fully separate the dry and wet contributions to the delay, which is a
more physically correct approximation.  It is based on an ideal model
atmosphere whose temperature is constant from the surface to the
inversion layer $h_1$, then decreases linearly with height at rate $W$
from $h_1$ to the tropopause height $h_2$, then is 
assumed to be constant above $h_2$.  This mapping function is designed
to be a semi-analytic approximation to the atmospheric delay
integral that retains an explicit temperature profile that can be
determined using meteorological measurements.  The mapping function is
expanded as a second-order polynomial in $Z_d$ and $Z_w$, plus the
largest third-order term.  It is nonlinear in $Z_d$ and $Z_w$.  It
also contains terms which couple $Z_d$ and $Z_w$, thus including terms
which arise from the bending of the electromagnetic wave path through the
atmosphere.  The functional form for the atmospheric delay in this
\cite{Lanyi1984} model is given by: 
\begin{equation}
\label{eq:Lanyi}
\mathcal{L}_{atm}=\frac{F(E)}{\sin{E}},
\end{equation}
where
\begin{multline}
\label{eq:FofE}
F(E) = F_d(E) Z_d + F_w(E) Z_w \\
+ \frac{F_{b1}(E) Z^2_d +2F_{b2}(E)
  Z_dZ_w + F_{b3}(E) Z^2_w}{\Delta} + \frac{F_{b4}(E) Z^3_d}{\Delta^2},
\end{multline}
where $Z_d$ = dry atmospheric zenith delay, $Z_w$ = wet
atmospheric zenith delay, $F_{bn}$ = n-th bending contributions to the
delay, $\Delta$ = dry atmospheric scale height = $\frac{kT_0}{mg_m}$,
$k$ = Boltzmann's constant, $T_0$ = daily average surface temperature,
$m$ = mean molecular mass of dry air, and $g_m$ = gravitational
acceleration of the center of gravity of the air column.  Standard
values of $k$, $m$, $T_0=292 K$ (appropriate for mid-latitudes), $g_m
= 978.37$ cm/s$^2$, and $\Delta=8.6$ km, are assumed.  The dry, wet,
and bending contributions are expressed in terms of moments of the
refractivity.  The bending terms are evaluated for the ideal model
atmosphere and thus give the dependence of the delay on the four
parameters $T_0$, $W$, $h_1$, and $h_2$.  Therefore, the
\cite{Lanyi1984} model relies upon accurate surface meteorological
measurements at the time of the observations to which the delay model
is applied.

Note that, contrary to the rest of this tutorial, we have cast the
functional form for the \cite{Lanyi1984} atmospheric delay in terms of
the elevation ($E$), which is the coordinate used by \cite{Lanyi1984},
rather than zenith angle.
As the terms in Equation~\ref{eq:FofE} are complex functions of $E$,
we opted not to provide a version of Equation~\ref{eq:FofE} which used
$z$ as the dependent variable, mainly out of fear of possibly adding
errors to this discussion.

\subsubsection{Mapping Function Summary}
\label{MappingSummary}

Differences between the various mapping functions increase rapidly at
high zenith angle ($z > 80^\circ$).  \cite{Lanyi1984} has compared the
\cite[][ Equation~\ref{eq:Marini}]{Marini1972},
\cite[][ Equation~\ref{eq:Chao}]{Chao1974}, and
\cite[][ Equation~\ref{eq:Lanyi}]{Lanyi1984} mapping functions for
atmospheric refractive delay measurements at radio wavelengths.  For
$z < 50^\circ$ these mapping functions differ by less than 4\,mm in
excess path length.  At high zenith angles ($z > 80^\circ$), though,
these differences increase to 60\,mm, rapidly increasing for even
higher zenith angles.  

Errors in the atmospheric path delay to
an antenna are equivalent to errors in the vertical position of the
antenna.  Furthermore, for an interferometric antenna array errors in
the vertical position of an antenna are to first-order proportional to
an error in the interferometric baseline involving that antenna.
Interferometric array baseline determination relies on measurements of
astronomical point sources observed over as large a range in zenith
angle as possible.  \cite{Davis1985} showed that limiting the
maximum zenith angle in a baseline measurement from 85 to 80 degrees
results in an error in the baseline determination of $\sim 10^{-5}$.
As baselines in an interferometric array must be measured to an accuracy
of better than one part in $10^7$ \citep{TMS2001} so as not to degrade
the sensitivity of the measurements made with the interferometric
array, errors in the determination of the atmospheric refractive delay
can be significant for Very Long Baseline Interferometric (VLBI)
measurements and/or interferometric measurements at millimeter and
submillimeter wavelengths.

\section{Some Background on Generator Function References}
\label{GenFuncReferences}

In the following we give some background information on some of the
references quoted in this section:
\begin{description}
\item[\cite{Niell1996}:] \textit{Global Mapping Functions for the
  Atmospheric Delay at Radio Wavelengths}.  The standard reference
  for the derivation of a global mapping function for atmospheric 
  delay.  This derivation of the mapping function is noteworthy in that
  it attempts to represent analytically the global 
  weather variations as a function of location (latitude) and time of
  year, and contains no adjustable parameters (\ie\ does not require
  input pressure and temperature for each station).  Note that
  Equation 4 in \cite{Niell1996} has a typo whereby the terms which are
  printed as ``1/term'' in both the numerator and denominator should
  really be just ``term'' in both the numerator and denominator.
\item[\cite{Davis1985}:] \textit{Geodesy by Radio Interferometry: Effects of
  Atmospheric Modeling Errors on Estimates of Baseline Length}.  An
  application of a modified Smith-Weintraub refractivity and the Niell
  mapping functions.
\item[\cite{Sovers1998}:] \textit{Astrometry and Geodesy with Radio
  Interferometry: Experiments, Models, Results}.  An excellent
  overview paper describing the details involved in calculating
  geometric and atmospheric delay.  Uses the \cite{Lanyi1984} model for
  the mapping function, which is a significant departure from the
  standard (\ie\ \cite{Niell1996}) mapping functions which derive from the
  \cite{Marini1972} reduced fraction functional form.
\item[\cite{Lanyi1984}:] \textit{Tropospheric Delay Effects in Radio
  Interferometry}.  Derivation of a new ``tropospheric'' (really
  atmospheric) mapping function which, unlike previous mapping
  functions, takes account of second and third order effects in the
  refractivity which are due to refractive bending.  This derivation
  of the mapping function is noteworthy in that it does not
  fully separate the dry and wet contributions to the delay, making it
  a physically more exact representation.  It is claimed to be more
  accurate than previous (\ie\ Niell) mapping functions for $z <
  86^\circ$, and the error due to the derived analytic form for the
  mapping function is estimated to be less than 0.02\% for $z <
  84^\circ$.
\item[\cite{Yan1995}:] \textit{The Generator Function Method of the
    Tropospheric Refraction Corrections}.  Another derivation of a new
  ``tropospheric'' (really atmospheric) mapping function.  A cousin to
  existing reduced-fraction expansions of the mapping function. 
\item[\cite{Yan1996}:] \textit{A New Expression for Astronomical
    Refraction}.  Related to the \cite{Yan1995} reference above,
  but applied to the refraction calculation problem.  Using the
  \cite{Yan1995} and \cite{Yan1996} references one can apply a unified
  formalism to both the atmosphere-induced refractive delay and
  bending problems.
\end{description}

\section{Conclusions}
\label{Conclusions}

Modern astronomical measurements often require sub-arcsecond position
accuracy.  For the simplified model atmospheres presented in this
tutorial, which assume a spherical structure with hydrostatic
equilibrium for the dry component (mainly the stratosphere) and
uniform mixing of the wet component (mainly the troposphere) of the
Earth's atmosphere, radio astronomical measurements with position
accuracy $\lesssim 1^{\prime\prime}$ at zenith angles $\lesssim
75^\circ$ are achievable.  Any of the functional forms for refractive
bending and delay which assume a spherical atmosphere are satisfactory
in this simplified scenario.  For measurements at zenith angles
$\gtrsim 75^\circ$, or for more realistic atmospheric conditions which
violate the simple scenario described above, or when higher positional
accuracy than $\sim 1^{\prime\prime}$ is required, more care needs to be
taken in the algorithm choice for atmospheric refractive bending and
delay.

For accurate calculation of the refractive electromagnetic wave
bending and propagation delay at an Earth-bound observatory, we
recommend the following:
\begin{enumerate}
\item \textsl{Refractive Bending Calculation:} Use the \cite{Auer2000}
  method (Equation~\ref{eq:refrac5}) with the procedure described in
  Section~\ref{Bending}.  The refractivity ($N(P,T)$) is derived from
  an atmospheric model such as \cite{Liebe1989} or
  \cite{Pardo2001}.
\item \textsl{Refractive Delay Calculation:} Use
  Equation~\ref{eq:DelayPlane} with refractivity derived from an
  atmospheric model.  The best of the mapping function solutions to
  $\mathcal{L}_{atm}$ is the \cite{Lanyi1984} algorithm
  (Equation~\ref{eq:Lanyi}), which appears to be quite accurate to
  zenith angles as high as $\sim 85^\circ$.
\end{enumerate}


\acknowledgments

JGM benefited greatly from discussions regarding the proper
calculation of atmospheric path delay with Darrel Emerson, Dick
Thompson, and Ed Fomalont.  The referee for this manuscript, Jim
Moran, provided invaluable advice which resulted in a greatly improved
manuscript.  The National Radio Astronomy Observatory is a facility of
the National Science Foundation operated under cooperative agreement
by Associated Universities, Inc.


\appendix
\section{Atmospheric Optical Refractivity}
\label{N0opt}

Refractivity in the optical is cast in a slightly different form than
that in the radio due to the fact that at optical wavelengths
dispersion is important, and color must always be taken into account.
\cite{Birch1993} (see also \cite{Livengood1999}) state that the
optical refractivity is given by the following:
\begin{equation}
N^{opt}_0 = N_{STP}\times N_{TP} - N_{RH}
\label{eq:nopt}
\end{equation}
where
\begin{eqnarray}
N_{STP} &=& 83.4305 + \frac{24062.94}{130-\lambda^{-2}} +
\frac{159.99}{38.9-\lambda^{-2}} \\
N_{TP} &=& \frac{P_d}{1.01325\times 10^{3}}\frac{(273.15 + 15)}{T}
\frac{\Bigl[1 + (3.25602 - 0.00972 T)P_d\times 10^{-6}\Bigr]}{1.00047} \\
N_{RH} &=& P_w\times(37.345 - 0.401\lambda^{-2})\times 10^{-3}
\label{eq:nopt-terms}
\end{eqnarray}
with $P_d$ and $P_w$ in hPa, $T$ in K, and $\lambda$ in
$\mu$m.  Note that we have ignored the small correction for an
increase in $CO_2$ concentration in Equation \ref{eq:nopt}.

\section{Acceleration Due to Gravity}
\label{gm}

The mean acceleration due to gravity ($g_m$) at the center of mass of
a vertical column of air above an observer is given by:
\begin{equation}
\label{eq:gm}
g_m = \frac{\int^\infty_0 dx \rho(x) g(x)}{\int^\infty_0 dx \rho(x)}.
\end{equation}
By expanding $g(x)$ to first-order in x, fits to harmonic
forms of $g_m$ as a function of latitude ($\phi$) can be derived.
Geodesists use a closed form of this harmonic function fit, known as
the Somigliana-Pizzetti formula
\citep{Pizzetti1894,Somigliana1929,Moritz1980}: 
\begin{eqnarray}
\label{eq:SPformula}
g_m &=& \frac{a\gamma_e\cos^2{\phi} +
  b\gamma_p\sin^2{\phi}}{\sqrt{a^2\cos^2{\phi}+b^2\sin^2{\phi}}}
\nonumber \\
&=& \gamma_e\frac{1+\kappa\sin^2{\phi}}{\sqrt{1+e^2\sin^2{\phi}}},
\end{eqnarray}
where $a$ and $b$ are the semimajor and semiminor axes of the
geocentric gravitational potential ellipsoid of revolution chosen to
define the Earth's gravitational potential, $\gamma_e$ and
$\gamma_p$ are the theoretical gravitational acceleration at the
Earth's equator and pole, respectively, and $e$, the first eccentricity
of the ellipsoid, and $\kappa$ are defined as follows:
\begin{eqnarray}
\label{eq:ekappa}
e \equiv \sqrt{1-\left(\frac{b}{a}\right)^2} \\
\kappa \equiv \frac{b\gamma_p}{a\gamma_e} - 1.
\end{eqnarray}
Two Chebyshev approximations to Equation~\ref{eq:SPformula} are in
common usage in geophysics.  The first has a relative accuracy of
$10^{-3}$\,$\mu$m sec$^{-2}$ \citep{Moritz1980} and is given by:
\begin{equation}
\label{eq:gmapprox1}
g_m = \gamma_e\left(1+\alpha_0\sin^2{\phi}+\alpha_1\sin^4{\phi}+\alpha_2\sin^6{\phi}+\alpha_3\sin^8{\phi}\right),
\end{equation}
while the second has a relative accuracy of 1\,$\mu$m sec$^{-2}$
\citep{Moritz1980} and is given by:
\begin{equation}
\label{eq:gmapprox2}
g_m = \gamma_e\left(\beta_0\sin^2{\phi}+\beta_1\sin^2{2\phi}\right).
\end{equation}
Equation~\ref{eq:gmapprox2} is the approximation most often used to
compute the latitudinal dependence of gravitational acceleration in
geophysics.

Most derivations of the mean acceleration due to gravity
at a given latitude calculate this quantity with
reference to the center of mass of a vertical column of air above an
observer ($H_c$).  It is often convenient to calculate $g_m$ as a 
function of the height of an observer above sea level on the surface
of the Earth ($h_0$).  \cite{Saastamoinen1972} points out that, due to
the poleward slope of the tropopause and seasonal variations of $T$
and $P$, regional and seasonal variations in $H_c$ tend to be smoothed
out.  To an accuracy of $\pm0.4$ km, $H_c$ and $h_0$ are related by:
\begin{equation}
\label{eq:hc}
H_c = 0.9~h_0 + 7.3~km
\end{equation}

In the following we list a variety of formulations for $g_m$ as
functions of latitude ($\phi$) and observer height above sea level
($h_0$, in km).  These expressions for $g_m$ differ by the assumed 
gravitational potential ellipsoid and, with the exception of
Equation~\ref{eq:gmwgs84}, rely on the use of the approximate form for
$g_m$ given in Equation~\ref{eq:gmapprox2}.  For observer height above sea
level ranging from 0 to 25\,m all of the expressions for $g_m$ listed
in this appendix differ by less than 0.015\%.  Any of the 
$g_m$ listed below are sufficient for the refraction application.
Note also that none of these expressions for $g_m$ take account of
local gravitational variations such as from nearby mountains, which can be
significant.

\begin{figure}
  \centering
  \includegraphics[scale=0.75]{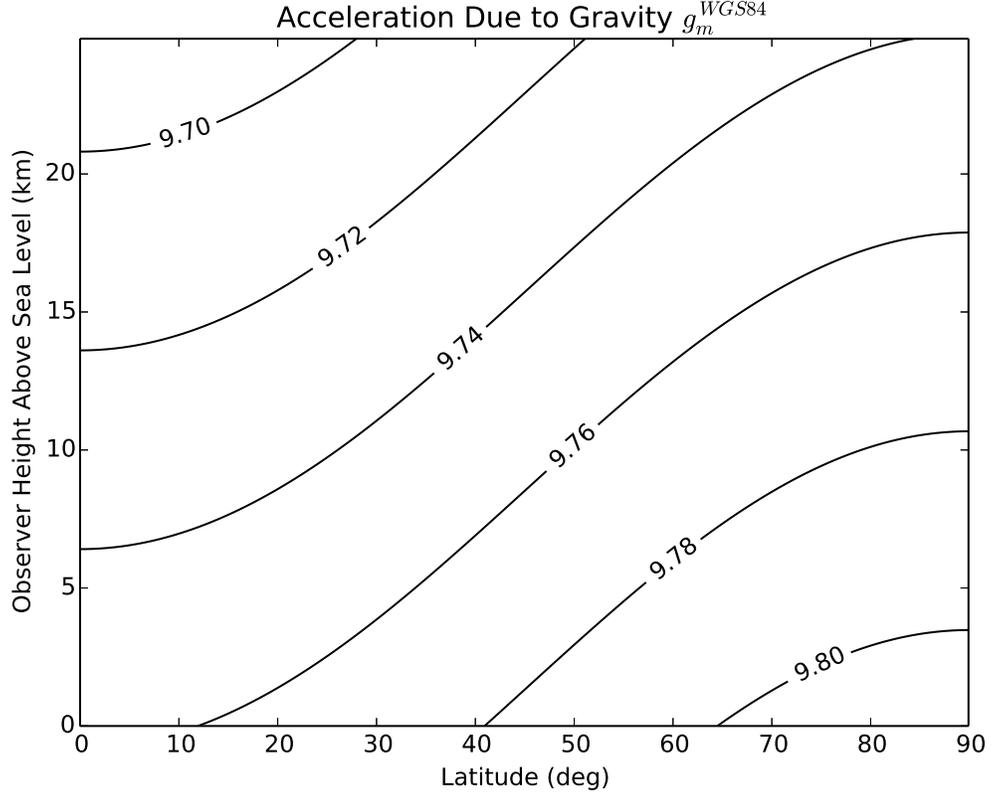} \\
  \caption{Acceleration due to gravity $g^{WGS84}_m$ as a function of
    observer latitude and height above sea level.}
  \label{fig:gmwgs84}
\end{figure}

The expression for $g_m$ that we have adopted in this work comes from
the definition adopted by the World Geodetic System 1984 (WGS84), with
an additional height correction: 
\begin{eqnarray}
\label{eq:gmwgs84}
g^{WGS84}_m &=& 9.7803267714\left(\frac{1 + 0.00193185138639
    \sin^2\phi}{\sqrt{1 - 0.00669437999013 \sin^2\phi}}\right) -
0.003086~H_c~m/s^2 \nonumber \\
&=& 9.7803267714\left(\frac{1 + 0.00193185138639
    \sin^2\phi}{\sqrt{1 - 0.00669437999013 \sin^2\phi}}\right) -
0.02253 \nonumber \\
&& - 0.0027774~h_0~m/s^2 
\end{eqnarray}
where $h_0$ is the height of the observer and $H_c$ is the
height of the center of mass of the vertical column of air above the
observer, both in km.  Figure~\ref{fig:gmwgs84} shows how
$g^{WGS84}_m$ varies as a function of latitude and observer height
above sea level.

\cite{Allen1973} quotes the following form:
\begin{eqnarray}
\label{eq:gmallen}
g^{Allen}_m &=& 9.80612 - 0.025865 \cos(2\phi) + 0.000058 \cos^2(2\phi) -
0.00308~H_c~m/s^2 \nonumber \\
           &=& 9.780313\left(1 + 0.005289 \sin^2\phi - 0.0000059
             \sin^2(2\phi) - 0.000315~H_c\right)~m/s^2 \nonumber \\
           &=& 9.757823\left(1 + 0.005301 \sin^2\phi - 0.0000059 
             \sin^2(2\phi) - 0.000284~h_0\right)~m/s^2 \hspace{10mm}
\end{eqnarray}
From \cite{Urban2013} (which is also the form used in
\textsl{SLALIB} and by \cite{Hohenkerk1985}, and where $h_0$ is in km):
\begin{equation}
\label{eq:gmes}
g^{ES}_m = 9.784 \left(1.0 - 0.0026 \cos(2\phi) -
  0.00028~h_0\right)~m/s^2
\end{equation}
The CRC handbook gives yet another variant:
\begin{eqnarray}
\label{eq:gmcrc}
g^{CRC}_m &=& 9.780356 \left(1 + 0.0052885 \sin^2\phi - 0.0000059
  \sin^2(2\phi)\right) - 0.003086~H_c~m/s^2 \nonumber \\
 &=& 9.757828 \left(1 + 0.005301 \sin^2\phi - 0.0000059
  \sin^2(2\phi) - 0.000284~h_0\right)~m/s^2 \hspace{10mm}
\end{eqnarray} 
with the reference \cite{Jursa1985}.  The web site
\begin{verbatim}http://geophysics.ou.edu/solid_earth/notes/potential/igf.htm
\end{verbatim}
lists the
following, which is based on the Geodetic Reference System 1967:
\begin{eqnarray}
\label{eq:gmigf67}
g^{IGF67}_m &=& 9.78031846 \left(1 + 0.0053024 \sin^2\phi - 0.0000058
  \sin^2(2\phi)\right) - 0.003086~H_c~m/s^2 \nonumber \\
 &=& 9.757791 \left(1 + 0.005315 \sin^2\phi - 0.0000058
  \sin^2(2\phi) - 0.000284~h_0\right)~m/s^2 \hspace{10mm}
\end{eqnarray}
where we have added the free-air and height correction
term.  Finally, \cite{Saastamoinen1972} derives:
\begin{eqnarray}
\label{eq:gmsaas}
g^{Saast}_m &=& 9.8062 \left(1 - 0.00265 \cos(2\phi) -
  0.00031~H_c\right) \nonumber \\
 &=& 9.784 \left(1 - 0.00266 \cos(2\phi) - 0.00028~h_0\right)
\end{eqnarray}

\section{Relative Humidity and Saturation Vapor Pressure}
\label{RHandPsat}

Note that the relative humidity at the observer ($RH_0$, in percent)
is related to the saturation vapor pressure ($e_{sat}$, in hPa;
\cite{Buck1981}) as follows (see \cite{Crane1976})
\begin{eqnarray}
\label{eq:rhesat}
e_{sat} &=& \left(1.0007 + 3.46\times10^{-6}
  P_0\right)6.1121\exp\Biggl[\frac{17.502 T_0}{T_0+240.97}\Biggr] \\
P_{w0} &=&
e_{sat}RH_0\left[1-\left(1-RH_0\right)\frac{e_{sat}}{P_0}\right]^{-1}
\end{eqnarray}
This relationship between $e_{sat}$, $P_{w0}$, and $RH_0$
is useful when using expressions for $N_0$ which involve linear
and quadratic expansions in $P_0$ and $P_{w0}$.  Unfortunately, this
complicated form for $e_{sat}$ does not yield itself to closed-form
integration.

By assuming that the relative humidity remains constant throughout the
troposphere, and equal to its value at the observer ($RH(r) = RH_0$),
we can write:
\begin{equation}
\label{eq:pw}
\frac{P_w}{P_{w0}} = \frac{e_{sat}(P,T)}{e_{sat}(P_0,T_0)}
\end{equation}
Tabulated values of $e_{sat}$ versus $T$ indicate that:
\begin{equation}
\label{eq:psatdep}
\frac{e_{sat}(P,T)}{e_{sat}(P_0,T_0)} = \left(\frac{T}{T_0}\right)^\gamma
\end{equation}
which yields:
\begin{equation}
\label{eq:psatdep2}
\frac{P_w}{P_{w0}} = \left(\frac{T}{T_0}\right)^\gamma
\end{equation}
As noted by \cite{Sinclair1982} and \cite{Hohenkerk1985},
the power index $\gamma$ is derived by fitting to the tabulated values
of $P_{sat}$ versus $T$ given in \cite{List1951}.  This fit
produces the following:
\begin{equation}
\label{eq:psatdep3}
P_{sat} = \left(\frac{T}{247.1}\right)^{18.36}
\end{equation}
Comparing this expression with that derived by
\cite{Buck1981} (Equation~\ref{eq:rhesat}) for P between 600\,hPa and
1200\,hPa and T between $-30$\,C and $+20$\,C indicates agreement to within
$\pm0.2$\,hPa.  Therefore, the approximate relation between $P_{sat}$
and $T$ (Equation~\ref{eq:psatdep3}) represents a good approximation
over this relevant range of $P$ and $T$.

\bibliographystyle{apj}
\bibliography{RefBendDelayCalcPaper}

\end{document}